\magnification 1200
\centerline {\bf  On the Mathematical Theory of Superfluidity}
\vskip 1cm
\centerline {\bf by Geoffrey L. Sewell* and Walter F. Wreszinski**}
\vskip 0.5cm
*Department of Physics, Queen Mary, University of London, Mile End Road, London E1 
4NS: e-mail g.l.sewell@qmul.ac.uk 
\vskip 0.5cm
**Instituto de Fisica, Universidade de Sao Paolo, CP 66318, 05315-970 Sao Paolo, 
Brazil: e-mail wreszins@gmail.com
\vskip 1cm
\centerline {\bf Abstract}
\vskip 0.3cm
We provide a general operator algebraic formulation of  the theory of superfluidity in 
Bose systems, with the aim of investigating the relationships of this phenomenon  both to 
off-diagonal long range order (ODLRO) and to a mathematically precise version of 
Landau\rq s picture of elementary excitations. Our principal results are that ODLRO 
leads both to rotational superfluidity and to Goldstone excitations, while the neo-Landau 
picture  accounts for the translational superfluidity of flow along a pipe. The latter picture 
is realised by the Lieb-Liniger-Girardeau model. Open problems are briefly discussed.
\vfill\eject
\centerline {\bf 1. Introduction}
\vskip 0.3cm
Superfluidity, or frictionless flow,  is a quantum phenomenon on the macroscopic scale, 
which arises in ordered phases of superconductors [Lo1], liquid HeII [2, 3] and 
trapped Bose gases [4]. The quantum theory of these phases is not only intrinsically 
fascinating but is also remarkable for its ramifications in other areas, such as the Higgs 
[5] mechanism in particle physics and the ordered structures, far from equilibrium, of 
laser light [6] and biocells [7]. Although there is an enormous literature on the subject, 
it is probably fair to say that there are two principal lines along which theories of 
superfluidity have developed over the years.
\vskip 0.2cm
The first of these is centred on the phenomenon of Bose-Einstein (BE) condensation. 
Specifically, Tisza [8] and London [2] proposed that the superfluid phase of HeII is 
characterised by a version of this condensation that prevails even in interacting Bose 
systems. The precise form of this condensation was subsequently formulated by O. 
Penrose and Onsager [9] and later extended to superconductors by Yang [10], who coined 
the term {\it off-diagonal long range order} (ODLRO) to describe it. In fact the 
characterisation of superconductivity by ODLRO encapsulated the essential content of  
the Bardeen-Cooper-Schrieffer (BCS) theory [11] . In general, the ODLRO condition is 
expressed in terms of a {\it macroscopic wave function}. For a system in a pure phase, 
this is just the position-dependent expectation value of a quantum field\footnote*{This 
description is appropriate if one describes the system in terms of its field algebra. While it 
is convenient to adopt this description, it is not really necessary, since one can equally 
well formulate ODLRO in terms of the observable algebra, given by the globally gauge 
invariant subalgebra of the field algebra}, which in the bosonic case is just that 
representing its particles, while in the fermionic case it is the product of the fields 
representing its particles of opposite spin. Thus, in either case, it corresponds to a 
spontaneous {\it gauge symmetry breakdown}.
\vskip 0.2cm
The second line of development of the theory of superfluidity stems from Landau\rq s 
[12] picture of the low lying excitations of HeII. This was based on a remarkable 
combination of quantum theory and hydrodynamics which depicted these excitations as 
phonons and rotons, corresponding to quantised sound waves and vortices, respectively. 
\vskip 0.2cm 
Although these two approaches to superfluidity are based on different ideas, they have 
been brought together, to some extent, by the works of Bogoliubov [13]
\footnote{**}{See also the excellent recent review of Bogoliubov\rq s work by 
Zagrebnov and Bru [14].} and Feynman [15] on interacting Bose systems (weakly 
interacting in Bogoliubov\rq s case). In fact, both of these works supported Landau\rq s 
picture and also satisfied the ODLRO condition.
\vskip 0.2cm
Further, both approaches led to models wherein a superfluid comprises two spatially 
coexisting fluids, the hydrodynamics of the first being irrotational and frictionless and 
that of the second being normal: thus the superfluid proper is just the first component. 
According to the BE condensation picture, this corresponds to the condensate, while in 
the Landau picture it corresponds to the residual ground state component of the mixed 
state of the system. In both pictures, the normal fluid component is carried by the 
excitations. At a phenomenological level, the two-fluid model has enjoyed great success 
in accounting for the observed thermodynamical properties of superfluids [1-3, 12].
\vskip 0.2cm
As regards the problem of superfluidity {\it per se}, the situation is less clear, at least on 
the level of mathematical physics. In the BE condensation, or ODLRO, picture, it 
depends on the {\it assumption} that the condensate flow is irrotational and frictionless, 
with velocity potential given by the phase angle of the macroscopic wave function. In 
Landau\rq s picture, the frictionlessness of the flow of  HeII at sufficiently small 
velocities stems from the stability of its ground state against the generation of quasi-
particle excitations, i.e. phonons and rotons. However, that picture carries no 
considerations of possible instabilities against more complicated excitations.  In fact, it 
appears to us that the existing rigorous results about superfluidity are confined to proofs 
that it prevails in (a) superconductors, subject to the assumptions of ODLRO, gauge 
covariance and thermodynamical stability [16], and (b) trapped, dilute, interacting Bose 
gases [4, 17, 18]. 
\vskip 0.2cm
The object of the present article is to make a general analysis of the phenomenon of 
superfluidity of suitably interacting bosonic systems, with the aim of obtaining sharp 
mathematical criteria for its occurrence. Here we emphasise that there are two different 
versions of this phenomenon. The first, which we shall call the {\it translational} one, 
concerns the frictionless flow of the superfluid along a cylindrical pipe: the second, 
which we shall call the {\it rotational} version, is the phenomenon whereby a superfluid 
in a rotating drum remains at rest and thus does not contribute to the moment of inertia of 
the drum-plus-fluid. In fact, we find that the conditions for these two manifestations of 
superfluidity are quite different from one another. The rotational kind will be shown to 
stem from a superselection rule associated with ODLRO, similar to that governing 
persistent currents induced by a trapped magnetic field in a superconducting ring. The 
translational kind, on the other hand, will be shown to be more complicated, in that 
further conditions need to be added to Landau\rq s quasi-particle picture in order to 
ensure the metastability of  frictionless flow along a pipe.   
\vskip 0.2cm
We formulate our treatment of superfluidity within the operator algebraic framework of 
quantum statistical physics, which is natural for the study of intrinsic properties of matter 
in the thermodynamic limit [19-22]. Thus we start, in Section 2, with a concise 
pedagogical formulation, within this framework, of the generic model of interacting Bose 
systems on which our treatment of superfluidity will be based; and in Section 3 we 
briefly review the conditions for spontaneous Galilei and gauge symmetry breakdown, 
with associated Goldstone bosonic excitations, in this model. We then pass on, in Section 
4, to a treatment of rotational superfluidity, showing that this stems from a superselection 
rule associated with ODLRO: this includes a derivation of the Onsager-Feynman [15, 23] 
quantisation rule for vortices in superfluids. Section 5 is 
devoted to a treatment of the problem of translational superfluidity. There we show that 
the frictionless flow along a pipe cannot be stable against {\it all} localised 
modifications of state. Evidently, this raises questions about the character of the 
metastability of  this flow, and these are addressed in Section 6, where a refined version 
of Landau\rq s picture is formulated in terms of the elementary  excitations in superfluids 
and explicitly realised by the Lieb-Liniger-Girardeau model [24, 25]. We conclude in 
Section 7 with a discussion of both the results and the open questions concerning the 
stability or metastability of superfluid states.
\vskip 0.2cm
Throughout the article we shall employ units in which ${\hbar}, \ k_{Boltzmann}$ and 
the mass per particle are unity.
\vskip 0.5cm
\centerline {\bf 2. The Model}
\vskip 0.3cm
We take our model to be an infinitely extended system, ${\Sigma}$, of bosons of one 
species that occupies a $d$-dimensional Euclidean space, $X$, and we formulate this 
model in standard operator algebraic terms (cf. [19-22]). 
\vskip 0.3cm
{\bf The Field Algebra.} This is a $C^{\star}$-algebra of the quantised field representing 
the particles of ${\Sigma}$. It is constructed as an algebra of operators in the Fock-
Hilbert space, ${\cal H}_{0}$, which together with its Fock vacuum vector, 
${\Phi}_{0}$,  is defined by the following conditions. 
\vskip 0.2cm\noindent
(i) There is a map $W$, the Weyl map, of  $L^{2}(X)$ into the unitaries in ${\cal 
H}_{0}$ that satisfies the canonical commutation relations (CCR)
$$W(f)W(g)=W(f+g){\rm exp}\bigl(iIm(f,g)\bigr) \ {\forall} \ 
f,g{\in}L^{2}(X),\eqno(2.1)$$
where $(.,.)$ is the $L^{2}(X)$ inner product.
\vskip 0.2cm\noindent
(ii) ${\Phi}_{0}$ is cyclic with respect to the algebra of the polynomials of these 
operators.
\vskip 0.2cm\noindent
(iii) The Fock vacuum state ${\phi}_{0}:=({\Phi}_{0},.{\Phi}_{0})$ is given by the 
formula  
$${\phi}_{0}\bigl(W(f)\bigr)=({\Phi}_{0},W(f){\Phi}_{0}) =
{\rm exp}\bigl(-{1\over 2}{\Vert}f{\Vert}^{2}\bigr) \ {\forall} \ 
f{\in}L^{2}(X).\eqno(2.2)$$
\vskip 0.2cm\noindent
It then follows from (i)-(iii) (cf. [20, Ch. 3]) that $W(f)$ takes the form
$$W(f)={\rm exp}\bigl[i\bigl({\psi}(f)+{\psi}(f)^{\star}\bigr)\bigr],\eqno(2.3)$$
 where ${\psi}(f)$ and ${\psi}(f)^{\star}$ are closed densely defined operators in 
${\cal H}_{0}$ with the properties that 
\vskip 0.2cm\noindent
(a) ${\psi}(f)$ annihilates ${\Phi}_{0}$, 
\vskip 0.2cm\noindent
(b) ${\Phi}_{0}$ is cyclic with respect to the algebra of the polynomials in 
${\lbrace}{\psi}(f)^{\star}{\vert}f{\in}L^{2}(X){\rbrace}$, and 
\vskip 0.2cm\noindent
(c) ${\psi}$ and ${\psi}^{\star}$ satisfy the following form of the CCR.
$$[{\psi}(f),{\psi}(g)^{\star}]=(g,f).\eqno(2.4)$$ 
Thus, ${\psi}(f)$ is a smeared version, $\int_{X}dx{\psi}(x)f(x)$, of a quantum field 
${\psi}(x)$, that satisfies the formal CCR
$$[{\psi}(x),{\psi}^{\star}(y)]={\delta}(x-y); 
[{\psi}(x),{\psi}(y)]=0.\eqno(2.4)^{\prime}$$
\vskip 0.2cm
In order to describe the local properties of the field, we introduce the set, ${\cal L}$, of 
bounded open subsets of $X$ and, for each ${\Lambda}$ in ${\cal L}$, we define ${\cal 
H}_{\Lambda}$ to be the subspace of ${\cal H}_{0}$ generated by the application to 
${\Phi}_{0}$ of the polynomials in 
${\lbrace}{\psi}(f)^{\star}{\vert}f{\in}L^{2}({\Lambda}){\rbrace}$ and 
${\cal F}_{\Lambda}$ to be the $W^{\star}$-algebra of bounded operators in this space. 
Thus,  the local algebras ${\lbrace}{\cal F}_{\Lambda}{\vert}{\Lambda}{\in}
{\cal L}{\rbrace}$ satisfy the conditions of isotony and local commutativity, namely
$${\cal F}_{\Lambda}{\subset}{\cal F}_{{\Lambda}^{\prime}} \ 
{\rm if} \ {\Lambda}{\subset}{\Lambda}^{\prime}$$ 
and 
$$[A,B]=0 \ {\forall} \ A{\in}{\cal F}_{\Lambda}, \ B{\in}
{\cal F}_{{\Lambda}^{\prime}} \ {\rm if} \ 
{\Lambda}{\cap}{\Lambda}^{\prime}={\emptyset},$$
respectively. In view of the isotony property, 
${\cal F}_{\cal L}:={\bigcup}_{{\Lambda}{\in}{\cal L}}{\cal F}_{\Lambda}$ is a 
$^{\star}$-algebra. Equipped with the ${\cal H}_{0}$ operator norm, it becomes  a 
normed 
$^{\star}$-algebra, whose completion, ${\cal F}$, is a $C^{\star}$-algebra, termed the 
quasi-local field algebra of the model.
\vskip 0.2cm
We define ${\gamma}, \ {\sigma}$ and ${\xi}$ to be the representations in $Aut({\cal 
F})$ of the additive groups $S^{(1)}, \ X$ and $X$, corresponding to gauge 
transformations, space translations and Galilei velocity boosts, respectively, by the 
formulae
$${\gamma}({\theta})[W(f)]=W\bigl(f{\rm exp}(i{\theta})\bigr) \ {\forall} \ 
{\theta}{\in}{\bf R}({\rm mod}2{\pi}),\eqno(2.5)$$
$${\sigma}(x)[W(f)]=W(f_{x}) \ {\forall}x{\in}X, \ {\rm with} \ 
f_{x}(y)=f(y-x).\eqno(2.6)$$
and
$${\xi}(v)[W(f)] =W(f_{v}) \ {\forall} \ v{\in}X, \ {\rm with} \ f_{v}(x)=
f(x){\rm exp}(iv.x);\eqno(2.7)$$
or, formally, 
$${\gamma}({\theta})[{\psi}(x)]={\psi}(x){\rm exp}(i{\theta}), \ 
{\sigma}(x)[{\psi}(y)]=
{\psi}(x+y) \ {\rm and} \ {\xi}(v)[{\psi}(x)]={\psi}(x){\rm exp}(iv.x).\eqno(2.8)$$
\vskip 0.3cm
{\bf The Algebra of Observables and its Automorphisms} Since observables are gauge 
invariant quantities, we take the $C^{\star}$-algebra, ${\cal A}$, of quasi-local bounded 
observables of ${\Sigma}$ to be the $C^{\star}$-subalgebra of ${\cal F}$ comprising its 
gauge invariant elements. Likewise, we define ${\cal A}_{\Lambda}$ and ${\cal 
A}_{\cal L}$ to be the algebras comprising the gauge invariant elements of ${\cal 
F}_{\Lambda}$ and ${\cal F}_{\cal L}$, respectively.
\vskip 0.2cm
Since, by Eqs. (2.5)-(2.7), ${\gamma}({\theta})$ commutes with both ${\sigma}(x)$ and 
${\xi}(v)$, it follows that ${\cal A}$ is stable under the latter two automorphisms. 
Consequently, their restrictions to ${\cal A}$ are automorphims of this algebra, and so 
${\sigma}$ and ${\xi}$ operate as representations of space translations and Galilei 
boosts, respectively, in $Aut({\cal A})$, as well as in $Aut({\cal F})$.  
\vskip 0.2cm
{\bf  Local Normality and Unbounded Observables.} We assume that, of the myriad 
mathematical states and representations of  ${\cal A}$, the physical ones are locally 
normal, i.e. that their restrictions to the local algebras ${\cal A}_{\Lambda}$ are normal, 
since this is precisely the condition that, with probability 1, they do not admit an infinite 
number of particles in any bounded region [26]. 
\vskip 0.2cm
Furthermore, the locally normal representations of ${\cal A}$, and also of ${\cal F}$, are 
those that reduce to the Fock representation in any bounded spatial region [26]. 
This ensures that the formulation of the bounded local observables can be canonically 
extended to the unbounded ones [27]. Specifically, those that are localised within 
${\Lambda}$ are represented by the unbounded self-adjoint operators affiliated to ${\cal 
A}_{\Lambda}$, i.e. by those that commute with the commutant of this algebra in ${\cal 
A}$. In particular, the observables $N_{\Lambda}, \ J_{\Lambda}$ and 
$T_{\Lambda}$, representing the number of particles, the current and the kinetic energy, 
respectively, in ${\Lambda}$ are given  in terms of a differentiable orthonormal basis,  
${\lbrace}f_{r}{\rbrace}$, of $L^{2}({\Lambda})$ by the equations (cf. [28])
$$N_{\Lambda}={\sum}_{r}{\psi}(f_{r})^{\star}{\psi}(f_{r}); \ J_{\Lambda}=
{i\over 2}{\sum}_{r}\bigl({\psi}(f_{r})^{\star}{\psi}({\nabla}f_{r})-
{\psi}({\nabla}f_{r})^{\star}{\psi}(f_{r})\bigr);$$ 
$$T_{\Lambda}=
{1\over 2}{\sum}_{r}{\psi}({\nabla}f_{r})^{\star}.{\psi}({\nabla}f_{r})
\eqno(2.9)$$ 
or formally,
$$N_{\Lambda}=\int_{\Lambda}dxn(x): \ J_{\Lambda}=
\int_{\Lambda}dxj(x); \ T_{\Lambda}=\int_{\Lambda}dxt(x),\eqno(2.9)^{\prime}$$
where 
$$n(x)={\psi}^{\star}(x){\psi}(x); \ j(x)=
-{i\over 2}\bigl({\psi}^{\star}(x){\nabla}{\psi}(x)-
[{\nabla}{\psi}^{\star}(x)]{\psi}(x)\bigr); \ t(x)=
{1\over 2}{\nabla}{\psi}^{\star}(x).{\nabla}{\psi}(x)\eqno(2.10)$$
and Neumann boundary conditions are employed (cf. [28]).
\vskip 0.3cm
{\bf Physical States, Representations and Dynamics.} The condition of local normality 
does not suffice for the characterisation of physical states and representations, as there are 
locally normal ones that do not support a $C^{\star}$-dynamics of ${\Sigma}$, as given 
by a natural limit of that of a finite version of this model [29-31]. Instead, the 
system admits just a $W^{\star}$-dynamics via such a limit, and this is confined to 
certain privileged representations, ${\pi}$, of ${\cal A}$ [31]. To be specific, the 
dynamics in an admissible representation ${\pi}$ corresponds to a one-parameter group, 
${\lbrace}{\alpha}(t){\vert}t{\in}{\bf R}{\rbrace}$, of automorphisms of ${\pi}({\cal 
A})^{{\prime}{\prime}}$, defined according to the following prescription [31]. For 
${\Lambda}{\in}{\cal L}$,  we assume that the Hamiltonian operator of  the finite 
version, ${\Sigma}_{\Lambda}$, of ${\Sigma}$ located in ${\Lambda}$ is an affiliate, 
$H_{\Lambda}$, of ${\cal A}_{\Lambda}$. The evolute of an observable $A \ 
({\in}{\cal A}_{\Lambda})$ of ${\Sigma}_{\Lambda}$ at time $t$ is then 
$A_{\Lambda}(t):=({\rm exp}(iH_{\Lambda}t)A{\rm exp}(-iH_{\Lambda}t)$ and our 
condition for a representation ${\pi}$ to be physical is that, for any $A{\in}{\cal 
A}_{\cal L},  \ {\pi}\bigl(A_{\Lambda}(t)\bigr)$ converges strongly to a limit, 
necessarily in ${\pi}({\cal A})^{{\prime}{\prime}}$, as a ${\Lambda}$ increases to 
$X$ in, say, Fisher\rq s [32] sense. This limiting procedure then serves to define the 
strongly continuous dynamical group ${\alpha}({\bf R}) \ ({\in}
Aut({\pi}({\cal A})^{{\prime}{\prime}})$ by the formula 
$${\alpha}(t)[{\pi}(A)]=s-{\rm lim}_{{\Lambda}{\uparrow}X}
{\pi}\bigl({\rm exp}(iH_{\Lambda}t)A{\rm exp}(-iH_{\Lambda}t)\bigr) \ {\forall} \ 
A{\in}{\cal A}, \ t{\in}{\bf R}.\eqno(2.11)$$
To lighten the notation, we define
$${\tilde A}:={\pi}(A) \ {\rm and} \ {\tilde A}_{t}:={\alpha}(t)[{\pi}(A)] \ {\forall} \ 
A{\in}{\cal A}, \ t{\in}{\bf R}.\eqno(2.12)$$
\vskip 0.2cm
We now assume that the physical representations of  ${\cal A}$ are the locally normal 
ones that support the dynamics given by this prescription. Correspondingly, we take the 
physical states of the model to be those whose GNS representations satisfy this 
physicality condition. Denoting the GNS triple of such a state ${\phi}$ by $({\cal 
H},{\pi},{\Phi})$, we define ${\tilde {\phi}}$ to be the canonical extension of ${\phi}$ 
to ${\pi}({\cal A})^{{\prime}{\prime}}$ according to the formula 
$${\tilde {\phi}}(M)=({\Phi},M{\Phi}) \ {\forall} \ M{\in}
{\pi}({\cal A})^{{\prime}{\prime}}.\eqno(2.13)$$
The state ${\phi}$ is then termed stationary if ${\tilde {\phi}}$ is invariant under 
${\alpha}({\bf R})$, in which case these automorphisms are unitarily implemented by a 
strongly continuous representation $U$  of  ${\bf R}$ in ${\cal H}$, as defined by the 
formula [33]
$$U(t){\tilde A}{\Phi}={\tilde A}_{t}{\Phi} \ {\forall} \ A{\in}{\cal A}, 
t{\in}{\bf R}.\eqno(2.14)$$ 
We denote by $H$ the Hamiltonian operator given by $-i$ time the generator of the 
group $U({\bf R})$, i.e.
$$U(t)={\rm exp}(iHt) \ {\forall} \ t{\in}{\bf R}.\eqno(2.15)$$  
Likewise, if ${\phi}$ is translationally invariant, the automorphisms ${\sigma}(X)$ are 
unitarily implemented  by a representation $S$ of $X$ in ${\cal H}$, as defined by the 
formula 
$$S(x){\pi}(A){\Phi}={\pi}\bigl({\sigma}(x)A\bigr){\Phi} \ {\forall} \ A{\in}{\cal A}, \ 
x{\in}X\eqno(2.16)$$
and, assuming that $S(x)$ is continuous in $X$, we denote by $P$ the momentum 
operator given by $-i$ time the generator of the group $S(X)$, i.e. 
$$S(x)={\exp}(iP.x) \ {\forall} \ x{\in}X.\eqno(2.17)$$
\vskip 0.2cm
{\bf Note.} This formulation of the states and dynamical automorphisms of the 
observable algebra ${\cal A}$ may readily employed for that of the states and dynamics 
of the field algebra ${\cal F}$. We note, in particular, that a state ${\phi}$ on ${\cal A}$ 
has a unique extension to a gauge invariant state ${\phi}_{\cal F}$ on ${\cal F}$, 
defined by the formula
$${\phi}_{\cal F}(F)={\phi}(pF), \ {\rm where} \  pF=(2{\pi})^{-1}\int_{0}^{2{\pi}}  
d{\theta}{\gamma}({\theta})F  \ {\forall} \ F{\in}{\cal F}.\eqno(2.18)$$
\vskip 0.3cm
{\bf The KMS Thermal Equilibrium Condition.} As first proposed by Haag, Hugenholtz 
and Winnink [34], the equilibrium states of a system are those that satisfy the Kubo-
Martin-Schwinger (KMS) condition. In the present setting, this condition, as applied to 
the state ${\phi}$ of ${\Sigma}$ at inverse temperature ${\beta}$, takes the following 
form.  For any pair of elements $A,B$ of ${\cal A}$, there is a function $F_{AB}$ on 
the strip 
${\rm Im}(z){\in}[0,{\beta}]$ of the complex plane, that is analytic in its interior and 
continuous on its boundaries, where it reduces to the forms
$$F_{AB}(t+i{\beta})={\tilde {\phi}}({\tilde A}_{t}{\tilde B})  \ 
{\rm and} \  F_{AB}(t)={\tilde {\phi}}({\tilde B}{\tilde A}_{t})  \ {\forall} \ t{\in}
{\bf R};\eqno(2.19)$$ 
or, formally,
$${\tilde {\phi}}({\tilde A}_{t}{\tilde B})=
{\tilde {\phi}}({\tilde B}{\tilde A}_{t+i{\beta}}) \ {\forall} \ A,B{\in}{\cal A}, \ 
t{\in}{\bf R},\eqno(2.19)^{\prime}$$ 
\vskip 0.2cm
Support for the general characterisation of thermal equilibrium by the KMS condition  is 
provided by the fact that
\vskip 0.2cm\noindent
(i) it implies the stationarity of the state concerned [35]; 
\vskip 0.2cm\noindent  
(ii) it is satisfied by the canonical equilibrium state of a finite system and by any infinite 
volume limit thereof; and
\vskip 0.2cm\noindent
(iii) it is equivalent to the various dynamical and thermodynamical equilibrium properties 
that are natural desiderata of equilibrium states [36-39].
\vskip 0.2cm\noindent
The set, ${\cal S}_{\beta}$, of KMS states on ${\cal A}$ at inverse temperature 
${\beta}$ is manifestly convex and, as proved by Emch et al [35], its extremal 
elements are primary and enjoy the essential properties of pure thermodynamical phases. 
Thus, the central decomposition of a KMS state serves precisely to resolve it into pure 
equilibrium phases. 
\vskip 0.3cm
{\bf Ground State Condition.} The condition for ${\phi}$ to be a ground state of 
${\Sigma}$ is that of KMS corresponding to ${\beta}={\infty}$ and reduces to the 
condition that the Fourier transform of the function, or distribution, $t{\rightarrow}{\tilde 
{\phi}}({\tilde B}{\tilde A}_{t})$ has support in ${\bf R}_{+}$. Equivalently, by Eqs. 
(2.14) and (2.15), it is equivalent to the condition that $H$ is a positive operator that 
annihilates the vector ${\Phi}$.  
\vskip 0.3cm	
{\bf  Advent of the Chemical Potential.} A remarkable work by Araki, Haag, Kastler and 
Takesaki [40] established that the thermodynamical parameter known as the 
chemical potential emerges from the structures of the algebras ${\cal A}$ and ${\cal F}$ 
and the KMS condition. Specifically, it established that, if ${\phi}$ is a primary element 
of  ${\cal S}_{\beta}$ , then its canonical gauge invariant extension to ${\cal F}$ 
satisfies the corresponding KMS condition with respect to the dynamical automorphism 
group ${\alpha}_{\mu}({\bf R})$, defined by the formula
$${\alpha}_{\mu}(t)={\alpha}(t){\gamma}(-{\mu}t),\eqno(2.20)$$
where ${\mu}$ is a real-valued parameter. Since, by Eqs. (2.5) and (2.9), 
$${\gamma}({\theta})A={\exp}(iN_{\Lambda}{\theta})A
{\exp}(-iN_{\Lambda}{\theta}) \ {\forall} \ A{\in}{\cal A}_{\Lambda}, \ {\theta}
{\in}[0,2{\pi}], \ {\Lambda}{\in}{\cal L},\eqno(2.21)$$ 
it follows from Eqs. (2.11) and (2.20) that the modification of ${\alpha}(t)$ by the factor 
${\gamma}(-{\mu}t)$ corresponds to the addition of the term $-{\mu}N_{\Lambda}$ to  
the local Hamiltonians $H_{\Lambda}$ and thus to the advent of a chemical potential of 
value ${\mu}$.
\vskip 0.3cm
{\bf Off-Diagonal Long Range Order.} The concept of long range order (ODLRO) is a 
generalisation to interacting systems of that of Bose-Einstein condensation. Specifically, 
the state ${\phi}$ is said to possess ODLRO [9, 10] if there is a complex classical field 
${\Psi}$ that does not tend to zero at infinity and that satisfies the condition 
$${\rm lim}_{{\vert}y{\vert}\to\infty}
\bigl[{\langle}{\phi};{\psi}^{\star}(x){\psi}(x^{\prime}+y){\rangle}-
{\overline{\Psi}}(x){\Psi}(x^{\prime}+y)\bigr]=0.\eqno(2.22)$$
In fact, this condition defines ${\Psi}$ up to a constant phase factor\footnote*{The proof 
of this assertion for pair fields in the fermion case [16; Prop. 3.1] carries through 
trivially for the present bosonic one.}. This field is termed the {\it macroscopic wave 
function} and represents the condensate. Correspondingly, the condensate density, 
${\rho}_{c}(x)$, and the associated current density, $j_{c}(x)$, are defined by the 
formulae
$${\rho}_{c}(x)={\overline {\Psi}}(x){\Psi}(x)\eqno(2.23)$$
and
$$j_{c}(x)=-{i\over 2}\bigl({\overline {\Psi}}(x){\nabla}{\Psi}(x)-
{\Psi}(x){\nabla}{\overline {\Psi}}(x)\bigr).\eqno(2.24)$$
We shall presently discuss the significance of ${\rho}_{c}$ and $j_{c}$, with respect to 
the phenomenon of superfluidity. 
\vskip 0.3cm 
{\bf General Phenomenological Picture: the Two-Fluid Model.} According to an 
empirically based hydrodynamical picture [2, 3, 12], a fluid in its condensed, superfluid 
phase behaves as a mixture of two fluids,  one of which flows frictionlessly and 
irrotationally, while the other enjoys normal classical viscous properties. These are 
termed the superfluid and normal components, respectively, of the fluid.Thus the 
position-dependent hydrodynamical density, ${\rho}(x)$, and current-density, $j(x)$, 
take the forms
$${\rho}(x)={\rho}_{s}(x)+{\rho}_{n}(x)\eqno(2.25)$$
and
$$j(x)={\rho}_{s}(x)v_{s}(x)+{\rho}_{n}(x)v_{n}(x),\eqno(2.26)$$
where ${\rho}_{s}$ and $v_{s}$ are the density  and drift velocity, respectively, of the 
superfluid component, while ${\rho}_{n}$ and $v_{n}$ are those of the normal 
component. Further, the irrotationality condition for the superfluid component is simply 
that
$${\rm curl}v_{s}(x)=0. \eqno(2.27)$$
\vskip 0.3cm
{\bf Relationship between Condensate and Superfluid Densities.} The available results 
on the relationship between the condensate density, ${\rho}_{c}$ and the superfluid 
density, ${\rho}_{s}$, are those of O. Penrose and Onsager [9] for HeII and of Lieb, 
Seiringer and Yngvason [18] for the trapped dilute Bose gas, discussed in the paragraph 
following Eq. (2.28). In the former case, it is  argued that ${\rho}_{c}$ is approximately 
8 per cent of ${\rho}_{s}$; while in the latter 
one, it is proved that ${\rho}_{s}={\rho}_{c}$ and that, at least in the case of uniform 
drift velocity,  the condensate and superfluid current densities are equal. 
\vskip 0.2cm
In view of these results, it is tempting to conjecture that, in general, ${\rho}_{s}$ and 
${\rho}_{s}v_{s}$ are simply  proportional to ${\rho}_{c}$ and $j_{c}$, respectively, 
with the same constant of proportionality. Assuming this to be the case, it follows from 
Eqs. (2.23) and (2.24) that
$$v_{s}={\nabla}\bigl({\rm arg}({\Psi})\bigr),\eqno(2.28)$$
and hence that the irrotationality condition (2.27) is satisfied.
\vskip 0.3cm
{\bf Note on the Lieb-Seiringer-Yngvason (LSY) Model} [4, 18]. This represents a 
trapped dilute Bose gas of $N$ particles in a three-dimensional cube of side L with 
periodic boundary conditions. The particles are assumed to interact via a two-body 
potential $V$, whose scattering length is $a$ and the diluteness of the gas is represented 
by the condition that  $Na/L$ remains fixed and finite in the limit 
$N{\rightarrow}{\infty}$. Thus the LSY model, as treated in this limit, is an infinite 
system, though one that is quite different from the model ${\Sigma}$ that we have just 
formulated. Lieb and Seiringer [17] have established that the ground state of the LSY 
model exhibits ODLRO and that its macroscopic wave-function ${\Psi}$ is determined 
by a variational principle, proposed by Gross [41] and Pitaevski [42]. Moreover, this 
result prevails in the situation where the gas is subjected to an external, one-body 
potential [43]. 
\vskip 0.5cm
\centerline {\bf 3. Breakdown of Galilei and Gauge Symmetries.} 
\vskip 0.3cm
Let us first consider the action of the Galilei boost ${\xi}(v)$ on the local current 
$J_{\Lambda}$. By Eqs. (2.8), (2.9)${\prime}$ and (2.10),
$${\xi}(v)J_{\Lambda}=J_{\Lambda}+N_{\Lambda}v,\eqno(3.1)$$
which implies that if ${\phi}(N_{\Lambda}){\neq}0$, then the state ${\phi}$ is not 
Galilei invariant. In other words, except in the trivial case when ${\phi}$ is the Fock 
vacuum, this state breaks the Galilei symmetry and hence [44] it supports excitations of 
Goldstone bosons corresponding to quantised density waves, i.e. 
phonons\footnote*{Strictly speaking, it is only at a formal level that this has been 
demonstrated for continuous systems in Refs. [44] and [45].}.
 \vskip 0.2cm
A different kind of symmetry breakdown and associated Goldstone bosons arises in the 
case where ${\phi}$ satisfies both the KMS and ODLRO conditions. The argument runs 
as follows. Since ${\phi}$ is a primary KMS state, it follows [40] that its gauge 
invariant extension ${\phi}_{\cal F}$ to the field algebra ${\cal F}$ satisfies the KMS 
condition w.r.t. the automorphisms ${\alpha}_{\mu}({\bf R})$ defined by Eq. (2.20). 
Furthermore [35], the central decomposition of ${\phi}_{\cal F}$ resolves this state 
into primary ${\alpha}_{\mu}$-KMS states on ${\cal F}$ and thus takes the form
$${\phi}_{\cal F}=\int_{\cal P}{\nu}dP({\nu}),\eqno(3.2)$$
where $P$ is a probability measure on this set of primaries, which we denote by 
${\cal P}$. Hence, as ${\phi}$ is the restriction of ${\phi}_{\cal F}$ to ${\cal A}$,
$${\phi}=
\int_{\cal P}{\nu}_{{\vert}{\cal A}}dP({\nu}).\eqno(3.3)$$
Moreover, since ${\cal A}$ is the gauge invariant subalgebra of ${\cal F}$, it follows 
from Eq. (2.20) that  the automorphisms ${\alpha}_{\mu}(t)$ reduce to ${\alpha}(t)$ on 
${\cal A}$ and hence that the states ${\nu}_{{\vert}{\cal A}}$ are KMS states of 
${\Sigma}$ for ${\nu}$ a.e. w.r.t. $P$. Therefore as ${\phi}$ is a primary and hence 
[35] an extremal KMS state on ${\cal A}$ , it follows from Eq. (3.3) that
$${\nu}_{{\vert}{\cal A}}={\phi} \ {\rm for} \ {\nu}  \ {\rm a.e.} \ 
{\rm w.r.t} \ P.\eqno(3.4)$$
Consequently, as ${\psi}^{\star}(x){\psi}(x^{\prime}+y)$ is affiliated to ${\cal A}$, 
$${\langle}{\phi};{\psi}^{\star}(x){\psi}(x^{\prime}+y){\rangle}=
{\langle}{\nu};{\psi}^{\star}(x){\psi}(x^{\prime}+y){\rangle} \ {\rm for} \ 
{\nu}  \ {\rm a.e.} \ {\rm w.r.t.} \ P.\eqno(3.5)$$
Moreover, since ${\nu}$ is primary and therefore strongly clustering [35, 46],
$${\rm lim}_{y\to\infty}
[{\langle}{\nu};{\psi}^{\star}(x){\psi}(x^{\prime}+y){\rangle}-
{\overline {{\langle}{\nu};{\psi}(x){\rangle}}}
{\langle}{\nu};{\psi}(x^{\prime}+y){\rangle}]=0$$
and consequently by Eq. (3.5) 
$${\rm lim}_{y\to\infty}
[{\langle}{\phi};{\psi}^{\star}(x){\psi}(x^{\prime}+y){\rangle}-
{\overline {{\langle}{\nu};{\psi}(x){\rangle}}}
{\langle}{\nu};{\psi}(x^{\prime}+y)
{\rangle}]=0 \ {\rm for} \ {\nu} \ {\rm a.e.} \ {\rm w.r.t.} \ P.\eqno(3.6)$$
On comparing this equation with the ODLRO condition (2.22) and recalling that the latter 
formula defines the macroscopic wave function ${\Psi}$ up to a constant phase angle, it 
follows that
$${\langle}{\nu};{\psi}(x){\rangle}={\Psi}(x){\rm exp}(ic_{\nu}) \ {\rm for} \ {\nu} \ 
{\rm a.e.} \ {\rm w.r.t.} \ P,\eqno(3.7)$$
where $c_{\nu}$ is a real-valued constant. Hence, by Eq. (2.8),  
${\gamma}({\theta})[{\psi}(x)]={\psi}(x){\rm exp}(i{\theta})$, and therefore the 
components ${\nu}$ of ${\phi}_{\cal F}$  break the gauge symmetry. This implies 
rigorous bounds on the rate of clustering [47].
\vskip 0.2cm
This argument does not apply to the zero temperature situation for two reasons. Firstly, 
the work of Araki et al [40] concerning the extension of a primary KMS state ${\phi}$ on 
${\cal A}$ to an ${\alpha}_{\mu}$- KMS state on ${\cal F}$  is not applicable here, 
since 
ground states do not breed the modular automorphisms on which the argument depends. 
Secondly, the argument of Emch et al [35] concerning the central decomposition of KMS 
states 
into extremal KMS states depends on the finiteness of the inverse temperature ${\beta}$ 
and so is not applicable to ground states. 
\vskip 0.2cm
In view of this situation, we relate ODLRO to symmetry breakdown in ground states by a 
strategy different from the one employed above for thermal states. Specifically, instead of 
deriving the symmetry breakdown from ODLRO, we proceed in the opposite, and more 
usual, direction (cf. [45, 48]). Thus, we assume that the model supports primary 
ground states ${\nu}$ on ${\cal F}$ with respect to the automorphisms 
${\alpha}_{\mu}({\bf R})$ that fulfill the gauge symmetry breaking condition analogous 
to Eq. (3.7), i.e.
$${\langle}{\nu};{\psi}(x){\rangle}={\Psi}_{1}(x),\eqno(3.8)$$ 
where ${\Psi}_{1}$ is a classical field that does not tend to zero at infinity. It then 
follows from the clustering property of primary states that
$${\rm lim}_{{\vert}y{\vert}\to\infty}
\bigl[{\langle}{\nu};{\psi}^{\star}(x){\psi}(x+y){\rangle}-
{\overline {\Psi}}_{1}(x){\Psi}_{1}(x+y)\bigr]=0.$$
Hence, as ${\psi}^{\star}(x){\psi}(x+y)$ is affiliated to ${\cal A}$, the restriction 
${\phi}$ of ${\nu}$ to this algebra satisfies the ODLRO condition
$${\rm lim}_{{\vert}y{\vert}\to\infty}
\bigl[{\langle}{\phi};{\psi}^{\star}(x){\psi}(x+y){\rangle}-
{\overline {\Psi}}_{1}(x){\Psi}_{1}(x+y)\bigr]=0.$$
Further, in view its gauge covariance, the model must support different 
symmetry-breaking states ${\nu}$ on ${\cal F}$ that modify the function ${\Psi}_{1}$ 
by  constant phase factors running from $0$ to $2{\pi}$. Thus, the assumption of the 
condition (3.8) leads to the same picture of  the connections between ODLRO, symmetry 
breakdown and quasi-particle excitations for ground states as for thermal ones. However, 
this leaves open the question of whether ${\Sigma}$ supports translationally invariant 
ground states on ${\cal A}$ that possess ODLRO but do not extend to gauge symmetry 
breaking ground states on ${\cal F}$.
\vskip 0.2cm
Assuming, however, that the ground state ${\phi}$ exhibits ODLRO and extends to 
gauge symmetry breaking ground states, w.r.t. ${\alpha}_{\mu}({\bf R})$, on ${\cal 
F}$, we may infer [45, 48] from the above argument  that ${\phi}_{\cal F}$ supports 
Goldstone bosonic excitations corresponding to the action of  the field ${\psi}$ or 
${\psi}^{\star}$  on the cyclic vector ${\Phi}$. These are evidently single particle 
excitations and are realised in Bogoliubov\rq s treatment of weakly interacting Bose 
gases [13]. Assuming that the state ${\phi}$ is translationally invariant, as in a fluid 
rather than a crystalline phase, these excitations are naturally described in terms of the 
Fourier transform $a(k)$ of ${\psi}(x)$ and the states of ${\Sigma}$ carry them  take the 
form ${\phi}\bigl(a(k)^{\star}(.)a(k)\bigr)$ and ${\phi}\bigl(a(k)(.)a(k)^{\star}\bigr)$. 
We note here that although these excitations look quite different from the phonons 
carried by the Galilei symmetry breakdown, there are heuristic indications [49, 50], of a 
spectral nature, that they are closely related to them. We also remark that Lieb, Seiringer 
and Yngvason [18]  have employed a different definition of spontaneous symmetry 
breakdown, expressed in terms of Bogoliubov quasi-averages [13, 14]; and on that basis 
they have established the spontaneous breakdown of gauge symmetry in the ground state 
of their model of a dilute, trapped bose gas, which we discussed briefly at the end of 
Section 2. It is an open problem whether and how their picture of symmety breakdown  is 
related to that presented above.
\vskip 0.5cm
\centerline {\bf 4. ODLRO and Rotational Superfluidity}
\vskip 0.3cm
We base our treatment of rotational superfluidity on the picture of ${\Sigma}$ relative to 
a frame of reference that rotates with uniform angular velocity about an axis Oz, say. 
This is the picture of ${\Sigma}$ in a rotating drum, as viewed from a a frame of 
reference in which the drum is at rest and in the idealisation wherein the boundaries of 
the drum are at infinity. We assume that the system is subjected to a conservative external 
field that stabilises it against the centrifugal force, even in the limit where the drum 
becomes infinite. 
\vskip 0.2cm
Thus, the finite volume Hamiltonians, ${\lbrace}H_{\Lambda}^{\prime}{\rbrace}$, 
which govern the dynamics of ${\Sigma}$ relative to the rotating reference frame 
according to the formula (2.11), are given by the modification of Eq. (2.30) due to the 
application of Coriolis and centrifugal forces and an external conservative field. Hence 
$H_{\Lambda}$ takes the following form.
$$H_{\Lambda}^{\prime}={1\over 2}\int_{\Lambda}dx\bigl(i{\nabla}{\psi}^{\star}(x)-
A(x){\psi}^{\star}(x)\bigr) 
\bigl(-i{\nabla}{\psi}(x)-A(x){\psi}(x)\bigr)+$$
$$\int_{\Lambda}dxU(x){\psi}^{\star}(x){\psi}(x)
+\int_{{\Lambda}^{2}}dxdy{\psi}^{\star}(x){\psi}^{\star}(y)V(x-y){\psi}(y){\psi}(x),
\eqno(4.1)$$
where
$$A(x)={\omega}{\times}x,\eqno(4.2)$$
${\omega}$ is a constant angular velocity directed along $Oz$ and the centrifugal
potential is absorbed into $U(x)$. It will be assumed that the 
potentials $U$ and $V$ are both lower bounded and invariant under rotations about 
$Oz$. The latter demand implies that these rotations form a dynamical symmetry group 
of the model. We define the representation, ${\rho}$, of this group in 
$Aut({\cal A})$ by the following formula, expressed in terms of the cylindrical 
coordinates $(r,{\vartheta},z)$ of an arbitrary point $x$. 
$${\rho}({\vartheta}){\psi}(r,{\vartheta}^{\prime},z)={\psi}(r,{\vartheta}+
{\vartheta}^{\prime},z) \ 
{\forall} \ {\vartheta}{\in}[0,2{\pi}].\eqno(4.3)$$
Then the axial symmetry condition for the state ${\phi}$ is that 
$${\phi}{\circ}{\rho}({\vartheta})={\phi} \ {\forall} \ 
{\vartheta}{\in}[0,2{\pi}].\eqno(4.4)$$
\vskip 0.2cm
Assume now that ${\phi}$ satisfies both the ODLRO and the axial symmetry conditions. 
Then, it follows from Eqs. (2.22), (4.3) and (4.4) that, if ${\Psi}$ serves as a macroscopic 
wave function for ${\phi}$, then so too does ${\rho}({\vartheta}){\Psi}$. Therefore 
since, by the remark following Eq. (2.22), the ODLRO condition defines ${\Psi}$ up to a 
constant phase factor, it follows from Eq. (4.3) that
$${\rho}({\vartheta}){\Psi}(r,{\vartheta},z)={\kappa}({\vartheta}){\Psi}(r,{\vartheta},z),
\eqno(4.5)$$
where ${\kappa}({\vartheta})$ is a scalar of unit modulus. Further, since, by Eq. (4.3), 
$${\rho}({\vartheta}){\rho}({\vartheta}^{\prime})=
{\rho}({\vartheta}+{\vartheta}^{\prime}) \ {\rm and} \ {\rho}(0)=I,$$
 it follows from Eq. (4.5) that
$${\kappa}({\vartheta}){\kappa}({\vartheta}^{\prime})=
{\kappa}({\vartheta}+{\vartheta}^{\prime}) \ {\rm and} \ {\kappa}(0)=1,$$
which signifies that  ${\kappa}$ is a one-dimensional representation of the circle, i.e. that 
$${\kappa}({\vartheta})={\rm exp}(in{\vartheta}),$$
for some integer $n$, and consequently, by Eq. (4.5), that ${\Psi}$ takes the form
$${\Psi}(r,{\vartheta},z)=f(r,z){\exp}(in{\vartheta}).\eqno(4.6)$$
\vskip 0.3cm
{\bf  Superselection Sectors.} It follows immediately from the above specifications that 
the quantum number $n$ specifies a superselection sector, in that the normal folia of 
states with different values of this parameter are mutually disjoint. The integral character 
of $n$ represents the Onsager-Feynman [15, 23] quantisation rule.
\vskip 0.3cm
{\bf  Equilibrium and Metastable States.}  Assume now that, under the prevailing 
thermodynamic conditions, the equilibrium state is unique and that it satisfies the 
ODLRO and axial symmetry conditions. Then it follows immediately that the quantum 
number $n$ is zero when ${\omega}=0$ for the following reasons. If ${\omega}=0$, 
it follows from Eqs. (4.1) and (4.2) that $H_{\Lambda}^{\prime}$ is invariant under the 
time reversal antiautomorphism ${\tau}$, which sends ${\psi}(x)$ to ${\psi}^{\star}(x)$. 
Hence, by Eqs. (2.11) and (2.19), the corresponding equilibrium state is ${\tau}$-
invariant and therefore, if ${\Psi}$ serves as the macroscopic wave function for this state 
then, by  the ODLRO condition (2.22), so too does ${\overline {\Psi}}$. In view of the 
remark following Eq. (2.22), this implies 
that ${\Psi}$ and its complex conjugate differ only by a constant phase factor.  
Consequently, it follows from Eq. (4.6) that the winding number $n$ must be zero in this 
case. This argument may easily be generalised to show that the winding number, $n$, of 
an equilibrium state is reversed if ${\omega}$ is replaced by 
$-{\omega}$ in Eq. (4.1), i.e. that, in an obvious notation,
$$n_{\omega}=-n_{-{\omega}}.\eqno(4.7)$$
\vskip 0.2cm
The question now arises whether $n$ remains zero for some non-zero values of 
${\omega}$: if so we would have a kind of London rigidity against rotations [1]. 
This would manifest itself if  $n$ remained zero either 
\vskip 0.2cm\noindent
(a) in  the equilibrium states corresponding to a certain range of values of ${\omega}$; or
\vskip 0.2cm\noindent
(b) in nonequilibrium metastable states that are stabilised by the above superselection 
rule, as in this case of persistent currents in superconducting rings (cf. [16]). 
\vskip 0.2cm\noindent
One can readily envisage either possibility arising through a combination of the property 
(4.7) and the integral character of the winding number.
\vskip 0.3cm
{\bf Superfluidity of Condensate.} Assume now that the quantum number $n$ is zero for 
the state ${\phi}$, whether this is an equilibrium or a metastable state. Then, in this case,  
it follows from Eq. (4.6) that 
$${\Psi}(x)=f(r,z).\eqno(4.8)$$
In order to see the hydrodynamical consequences of this formula, we note that the 
presence of the vector potential $A$ in Eq. (4.1) for the formal Hamiltonian leads to an 
additional term $-A{\psi}^{\star}{\psi}$ in the formula for the current density observable 
in Eq. (2.10). Correspondingly it leads to a modification of the formula (2.24) for the 
condensate current density by the addition of the term $-A{\vert}{\Psi}{\vert}^{2}$, 
while leaving Eq. (2.23) for the condensate density unchanged. Thus, 
using Eq. (4.2), we see that the formula (2.24) is changed to 
$$j_{c}(x)=-{i\over 2}\bigl({\overline {\Psi}}(x){\nabla}{\Psi}(x)-
{\Psi}(x){\nabla}{\overline {\Psi}}(x)\bigr)-
{\vert}{\Psi}(x){\vert}^{2}({\omega}{\times}x).\eqno(4.9)$$
Hence, by Eqs.(2.23), (4.8) and (4.9), the transverse component of the condensate current 
is 
$$j_{c}^{tr}(x)=-{\rho}_{c}(x)({\omega}{\times}x)=-
{\vert}f(r,z){\vert}^{2}({\omega}{\times}x).\eqno(4.10)$$
This signifies that the transverse component of the condensate drift velocity, as viewed in 
the rotating frame, is $-{\omega}{\times}x$ and therefore that it is zero from the 
standpoint of an observer in the rest frame. This is just the manifestation of rotational 
superfluidity, wherein the condensate remains at rest while the container rotates [2]. 
Here we remark that this interpretation is dependent on the assumption of Section 2 
(following Eq. (2.27)) that the condensate and superfluid drift velocities are the same.    
\vskip 0.3cm
{\bf Comment.} The general argument presented here is supported by results on 
rotational superfluidity obtained by Lewis and Pule [51] for the ideal Bose gas below its 
transition temperature and by Lieb and Seiringer [52] for the dilute interacting Bose gas 
at  zero temperature. Note that, in the latter work, spontaneous symmetry breakdown is 
related to the appearance of two or more vortices. These are massive, i.e. not gapless, 
excitations. Indeed, the proofs of Goldstone\rq s theorem in Refs. [44] and [45] are not 
applicable here. The latter  depends on the assumption of translational invariance and 
the former relies on an unproved assumption concerning the fall-off  of 
${\Vert}[{\sigma}(x)\alpha(t)A,B]{\Vert}$, with increasing ${\vert}x{\vert}$ for fixed t 
and arbitrary local 
bounded observables $A$ and $B$.
\vskip 0.5cm
\centerline {\bf  5. Translational Superfluidity and the Local Instability}
\vskip 0.2cm
\centerline {\bf of Current-Carrying States}
\vskip 0.3cm
Translational superfluidity corresponds typically to frictionless flow along a pipe and 
may be represented by translationally invariant current-carrying states of the model 
${\Sigma}$ in the idealisation where the boundaries of the pipe are at infinity. Evidently, 
such states cannot enjoy global thermodynamical stability, since it follows from Galilei 
covariance that their free energy densities may be reduced by boosts opposing their drift 
velocities. It follows that the observed frictionless flow of superfluids must be carried by 
metastable, rather than globally stable, states. One may reasonably ask whether their 
metastability amounts to thermodynamical stability against modifications of state that are 
confined to bounded spatial regions, granted that there are models of other systems that 
support metastable states characterised by such local stability [39]. However, as we shall 
now show, translationally invariant current-carrying states of the present model, 
${\Sigma}$, do not enjoy that enjoy that kind of stability and therefore a different picture 
of their metastability is required. Such a picture will be proposed in Section 6. 
\vskip 0.2cm
The objective of this Section, then, is to show that a translationally invariant, current-
carrying, locally normal state, ${\phi}$, of the model ${\Sigma}$ cannot be energetically 
stable  against strictly localised modifications: in other words, for any such state 
${\phi}$, there is another state ${\phi}^{\prime}$ that coincides with ${\phi}$ outside 
some bounded spatial region ${\Lambda}$ and whose energy is lower than that of 
${\phi}$. This signifies that ${\phi}$ cannot be locally thermodynamically stable (LTS), 
in the sense defined in Refs. [38,  39], at zero temperature.
\vskip 0.3cm 
{\bf 5.1. The Strategy.} We assume that ${\phi}$ is a translationally invariant, locally 
normal state of the model ${\Sigma}$ of Section 2, and that its drift velocity is $v \ 
({\neq}0)$. In order to construct a local modification of this state that lowers its energy, 
we start by introducing spheres ${\Gamma}, \ {\Gamma}_{1}, \ {\Gamma}_{2}$ and 
${\Gamma}_{3}$ that are centred at the origin and whose radii are $R, \ (R+a), \ (R+b), \ 
(R+c)$, respectively, with $c>b>a>0$.  Here, $a, \ b, \ c$ are fixed lengths, whereas $R$ 
is a variable parameter, which may be made arbitrarily large. Thus, the regions between 
these spheres may be regarded as \lq shells\rq\ and, of these, we denote  
${\Gamma}_{3}{\backslash}{\Gamma}$ and 
${\Gamma}_{2}{\backslash}{\Gamma}_{1}$ by ${\Theta}$ and ${\Theta}_{1}$, 
respectively. We construct a linear, identity-preserving transformation ${\tau}$ of ${\cal 
A}$, that is completely positive (CP) in the sense defined by Stinespring [53], and an 
automorphism ${\chi}$ of this algebra according to the following specifications. ${\tau}$ 
reduces to the identity map outside the shell ${\Theta}$ and, within this region,  its action  
on the position-dependent number density observable $n(x):={\psi}^{\star}(x){\psi}(x)$ 
serves to multiply it by a factor $g(x)^{2}$, where $g$ is a smooth  function of $X$ into 
$[0,1]$ that vanishes in ${\Theta}_{1}$. This action therefore serves to evacuate the 
shell ${\Theta}_{1}$. On 
the other hand, ${\chi}$ is a {\it local} gauge automorphism of ${\cal A}$ which 
rephases the field operator ${\psi}(x)$ by a factor ${\rm exp}\bigl(ih(x)\bigr)$, where 
$h$ is a smoothly varying function of position that vanishes outside ${\Gamma}_{2}$ 
and takes the form $-v.x$ in ${\Gamma}_{1}$. Thus, ${\chi}$ induces a position 
dependent boost, whose velocity is $-v$ in ${\Gamma}_{1}$ and zero outside 
${\Gamma}_{2}$. The form of $h$, and thus of the boost velocity ${\nabla}h$, in 
${\Theta}_{1}$ is irrelevant for our purposes, in view of the evacuation of this shell.
\vskip 0.2cm
We define 
$${\phi}^{\prime}:={\chi}^{\star}{\tau}^{\star}{\phi},\eqno(5.1)$$
where ${\tau}^{\star}$ and ${\chi}^{\star}$ are the duals of ${\tau}$ and ${\chi}$, 
respectively. Thus, by the above specifications, the action of ${\tau}^{\star}$ on 
${\phi}$ evacuates the shell ${\Theta}_{1}$ and the subsequent action of 
${\chi}^{\star}$  neutralises the drift in ${\Gamma}_{1}$. We show that, under very 
general conditions on ${\phi}$ and the two-body potential $V$, the net effect of these 
transformations then leads to an energy decrease  of order $R^{d}$ in ${\Gamma}$ and 
an energy change of order $R^{d-1}$ in the shell ${\Theta}$. Hence, for sufficienly 
large $R$, the transition ${\phi}{\rightarrow}{\phi}^{\prime}$ leads to a net decrease of 
energy. This establishes that ${\phi}$ is not locally stable at zero temperature. 
\vskip 0.3cm
{\bf 5.2. Constructions.} We construct the CP map ${\tau}$ and the automorphism 
${\chi}$ as locally normal transformations of ${\cal A}$. In accordance with the above 
specifications, we define ${\chi}$ to be  the restriction to ${\cal A}$ of the 
automorphism of ${\cal F}$ denoted by the same symbol and  formally defined by the 
equation
$${\chi}{\psi}(x)={\psi}(x){\rm exp}\bigl(ih(x)\bigr),\eqno(5.2)$$
where $h$ is a smooth, real-valued function on $X$ that vanishes outside 
${\Gamma}_{2}$ and takes the form $h(x)=-v.x$ in ${\Gamma}_{1}$. 
\vskip 0.2cm
We then define ${\tau}$ to be the composite of an isomorphism, ${\iota}$, of ${\cal F}$ 
into ${\cal F}{\otimes}{\cal F}$ and a projection, ${\pi}$, of this product algebra onto 
${\cal A}$, i.e. 
$${\tau}:={\pi}{\circ}{\iota},\eqno(5.3)$$ 
where ${\pi}$ and ${\iota}$ are defined by the formulae 
$${\pi}(F{\otimes}F_{0})={\phi}_{0}(F_{0})(pF) \ {\forall} \ 
F, F_{0}{\in}{\cal F},\eqno(5.4)$$
and
$${\iota}{\psi}(x)=g(x){\psi}(x){\otimes}I+
I{\otimes}\bigl(1-g(x)^{2}\bigr)^{1/2}{\psi}(x).\eqno(5.5)$$
Here $p$ is the projection of ${\cal F}$ onto ${\cal A}$, defined in Eq.(2.18),  
${\phi}_{0}$ is the Fock vacuum state defined by Eq. (2.2) and $g$ is a smooth, real-
valued function on $X$ such that 
\vskip 0.2cm\noindent
(a) $1{\geq}g(x){\geq}0 \ {\forall} \ x{\in}X$;
\vskip 0.2cm\noindent
(b) $g$ vanishes in the shell ${\Theta}_{1}$; 
\vskip 0.2cm\noindent
(c) $g$ takes the value unity in both ${\Gamma}$ and $X{\backslash}{\Gamma}_{3}$; 
and
\vskip 0.2cm\noindent
(d) ${\vert}{\nabla}g{\vert}$ has a finite, $R$-independent upper bound.   
\vskip 0.2cm\noindent
It follows easily from these specifications that ${\tau}$ is indeed a completely positive 
contractive transformation of ${\cal A}$ that reduces to the identity map outside the 
region ${\Gamma}{\cup}\bigl(X{\backslash}{\Gamma}_{3}\bigr)$.
\vskip 0.3cm
{\bf  The Energy Increment $E({\phi}{\vert}{\phi}^{\prime})$.} This is the increment in 
the energy of ${\Sigma}$ for a transition from ${\phi}$ to ${\phi}^{\prime}$. We 
formulate it in terms  of the local energy observable $H_{\Lambda}$ given by Eq. (2.30), 
which we rewrite in the following form.
$$H_{\Lambda}=\int_{\Lambda}dxt(x)+
\int_{\Lambda}^{2}dxdyn^{(2)}(x,y)V(x-y),\eqno(5.6)$$
where $t(x)$ is the kinetic energy density defined in Eq. (2.10) and $n^{(2)}(x,y)$ is the 
pair density given by the formula
$$n^{(2)}(x,y)={\psi}^{\star}(x){\psi}^{\star}(y){\psi}(y){\psi}(x).\eqno(5.7)$$
We shall assume that the two-body potential $V$ is positive and measurable.
\vskip 0.2cm 
The energy increment ${\Delta}E({\phi}{\vert}{\phi}^{\prime})$ is defined (cf. [38, 
39]) by the formula
$${\Delta}E({\phi}{\vert}{\phi}^{\prime})={\rm lim}_{{\Lambda}{\uparrow}X}
\bigl[{\phi}^{\prime}(H_{\Lambda})-{\phi}(H_{\Lambda})\bigr].\eqno(5.8)$$
Now, by Eqs. (5.1), (5.6) and (5.7), 
$${\phi}^{\prime}(H_{\Lambda})-{\phi}(H_{\Lambda})=
\int_{\Lambda}dx{\langle}{\phi};({\tau}{\chi}-I)t(x){\rangle}+
\int_{{\Lambda}^{2}}dxdyV(x-y)
{\langle}{\phi};({\tau}{\chi}-I)n^{(2)}(x,y){\rangle}.\eqno(5.9)$$
Further, it follows from Eqs. (2.2), (2.10), (5.2)-(5.5) and (5.7), together with the fact that 
${\psi}$ annihilates the Fock vacuum vector ${\Phi}_{0}$, that
$${\tau}{\chi}t(x)={\gamma}(x)^{2}\bigl(t(x)+j(x).{\nabla}(x)+$$
$${1\over 2}n(x)[{\nabla}h(x)]^{2}\bigr)+{1\over 2}[{\nabla}g(x)]^{2}n(x)+
{1\over 2}g(x){\nabla}g(x).{\nabla}n(x)\eqno(5.10)$$
and
$${\tau}{\chi}n^{(2)}(x,y)=g(x)^{2}g(y)^{2}n^{(2)}(x,y).\eqno(5.11)$$
Moreover, under the assumption that ${\phi}$ is a translationally invariant state carrying 
a current of drift velocity $v$, the expectation values of $n(x), \ j(x)$ and $t(x)$ for this 
state are constants ${\overline n}, \ {\overline n}v$ and ${\overline t}$, respectively, 
which we assume to be finite; while that of $n^{(2)}(x,y)$ is a non-negative valued 
function ${\overline n}^{(2)}(x-y)$ of the difference of its arguments. We shall assume 
that this latter function is measurable. It follows then from Eqs. (5.9)-(5.11) that 
$${\phi}^{\prime}(H_{\Lambda})-{\phi}(H_{\Lambda})=
\int_{\Lambda}dx\bigl[{\overline t}\bigl(g(x)^{2}-1\bigr)+{\overline n}g(x)^{2}
\bigl(v.{\nabla}h(x)+{1\over 2}({\nabla}h(x))^{2}\bigr)+
{1\over 2}{\overline n}\bigl({\nabla}g(x)\bigr)^{2}\bigr]+$$ 
$$\int_{{\Lambda}^{2}}dxdy\bigl(g(x)^{2}g(y)^{2}-1\bigr)
{\overline n}^{(2)}(x-y)V(x-y).$$
Recalling now that  the restrictions of $g$ to ${\Gamma}{\cup} 
(X{\backslash}{\Gamma}_{3})$ and ${\Theta}_{1}$ are $1$ and $0$, respectively, and 
that those of ${\phi}$ to ${\Gamma}_{1}$ and $X{\backslash}{\Gamma}_{2}$ are $-
v.x$ and $0$, respectively, it follows from this last equation that, for sufficiently large 
${\Lambda}$,  
$${\phi}^{\prime}(H_{\Lambda})-{\phi}(H_{\Lambda})=
-{1\over 2}{\overline n}v^{2}\int_{{\Gamma}_{1}}dxg(x)^{2}+
\int_{\Theta}dx\bigl[{\overline t}\bigl(g(x)^{2}-1\bigr)
+{1\over 2}{\overline n}\bigl({\nabla}g(x)\bigr)^{2}\bigr]+$$
$$\int_{{\Lambda}^{2}}dxdy
[g(x)^{2}g(y)^{2}-1]{\overline n}^{(2)}(x-y)V(x-y)$$
Hence, by Eq. (5.8) and the positivity and measurability of ${\overline n}^{(2)}$ and 
$V$,
$${\Delta}E({\phi}{\vert}{\phi}^{\prime})=
-{1\over 2}{\overline n}v^{2}\int_{{\Gamma}_{1}}dxg(x)^{2}+
\int_{\Theta}dx\bigl[{\overline t}\bigl(g(x)^{2}-1\bigr)
+{1\over 2}{\overline n}\bigl({\nabla}g(x)\bigr)^{2}\bigr]+$$
$$\int_{X^{2}}dxdy[g(x)^{2}g(y)^{2}-1]{\overline n}^{(2)}(x-y)V(x-y).\eqno(5.12)$$
\vskip 0.3cm
{\bf Theorem 5.1.} {\it Under the above assumptions, 
${\Delta}E({\phi}{\vert}{\phi}^{\prime})$ is negative for sufficiently large $R$. Hence, 
as ${\phi}^{\prime}$ coincides with ${\phi}$ outside the bounded region 
${\Gamma}_{3}$, the latter state is unstable against some local modifications thereof. }
\vskip 0.3cm
{\bf Proof .} Since $g$ takes the value $1$ in ${\Gamma}$ and lies in the range $[0,1]$ 
in the shell ${\Gamma}_{1}{\backslash}{\Gamma}$, the first term on the r.h.s. of Eq. 
5.12) is negative and $O(R^{d})$. On the other hand, as the volume of ${\Theta}$ is 
$O(R^{d-1})$, it follows from our specifications of $g$ and $h$ that the second term is 
$O(R^{d-1})$, while the third term is negative. This establishes that 
${\Delta}E({\phi}{\vert}{\phi}^{\prime})<0$ for $R$ sufficiently large.
\vskip 0.3cm
{\bf Comments.} (1) This theorem does not conflict with Landau\rq s picture of 
superfluidity [12], which requires the stability of a current carrying state against the 
excitation of certain quasi-particles (phonons and rotons), while making no explicit 
demand of its stability against all local modifications.
\vskip 0.2cm
(2) The theorem might seem to indicate that ${\phi}$ is not a ground state, in the strict 
sense specified in Section 2. However, attempts to prove this point are impeded by 
problems connected with the domains of unbounded operators of the model. 
\vskip 0.2cm
(3) Similarly, such problems impede attempts to prove that current carrying, 
translationally invariant states at finite temperatures cannot be thermal equilibrium states. 
\vskip 0.2cm
(4) On the other hand, in the case of lattice models, the argument employed to prove 
Theorem 5.1 has been extended [31] to show that translationally invariant, current 
carrying 
states cannot satisfy the equilibrium conditions either at zero or non-zero temperature.  
\vskip 0.5cm
\centerline {\bf 6. Generalised Landau States }
\vskip 0.3cm
{\bf 6.1. Stability of Current-Carrying States against Elementary Excitations.}
We have seen in the previous section that current-carrying states cannot be locally 
thermodynamically stable (Theorem 5.1). We are thus led to look for a weaker stability 
condition which is able to account for their observed metastability. The chosen condition 
pertains to elementary excitations, which, as we shall see, comprise a mathematically 
precise modification of Landau's quasi-particle picture. The basic heuristic idea behind 
this choice is that the interaction of the system ${\Sigma}$ and its environment is 
presumably of the few-particle type and thus the only kinetic processes it may be 
expected to generate are ones involving the creation of but a few elementary excitations, 
rather than the highly complicated ones considered in Section 5. 
\vskip 0.2cm
This Section is devoted to a treatment of the stability of current-carrying states against the 
elementary excitations. We start by introducing the Galilei boosted version of a ground 
state ${\phi}$, namely
$${\phi}_{v}={\phi}{\circ}{\xi}(v),\eqno(6.1)$$
where ${\xi}(v)$ is the boost given by Eq. (2.7); and we term ${\phi}_{v}$ a {\it 
generalised Landau state} if it is stable against elementary excitations, in a sense that will 
be made precise in due course (in Def. 6.1).
\vskip 0.2cm
Assuming the uniqueness of the ground state, ${\phi}$ and ${\phi}_{v}$ may be 
expressed as limits of their finite system counterparts in the following way. We define 
${\Lambda}_{L}$ to be the periodicised cube whose centroid is 
the origin and whose sides are parallel to the coordinate axes and are of length $L$; and 
we define $H_{N,L}$ to be the Hamiltonian of the $N$-particle version, 
${\Sigma}_{N,L}$, of ${\Sigma}$ that lives in 
this cube. Thus, $H_{N,L}$ is an operator in the $N$-particle subspace, 
${\cal H}_{N,L}$, of the Fock space, ${\cal H}_{L}$, over ${\Lambda}_{L}$. We 
denote by ${\Phi}_{N,L}$ its ground state vector. This is then a vector in ${\cal 
H}_{N,L}$, and thus in  ${\cal H}_{L}$, and hence the ground state  
${\phi}_{N,L}:=({\Phi}_{N,L},.{\Phi}_{N,L}$ of ${\Sigma}_{N,L}$ extends to the 
gauge invariant subalgebra, ${\cal A}_{L}$, of  ${\cal B}({\cal H}_{L})$ according to 
the formula
$${\phi}_{N,L}(A)=({\Phi}_{N,L},A{\Phi}_{N,L})  \ {\forall} \ 
A{\in}{\cal A}_{L}.$$
Note that this state annihilates the observables whose particle numbers are different from 
$N$. 
Further, the dynamics of ${\Sigma}_{N,L}$ is given by the automorphisms 
${\alpha}_{N,L}(t)$ 
of  ${\cal A}_{L}$ given by the formula
$${\alpha}_{N,L}(t)(A)={\rm exp}(iH_{N,L}t)A{\rm exp}(-iH_{N,L}t).\eqno(6.2)$$
Now the limits that concern us are those in which $N$ and $L$ tend to infinity in such a 
way that $N/L^{d}$ is fixed at a finite value ${\rho}$. We assume that this limiting 
procedure yields the 
ground state ${\phi}$ of ${\Sigma}$ and the dynamical automorphism group ${\alpha}$ 
of this system, according to the formulae
$${\phi}(A)={\rm lim}_{L\to\infty;N={\rho}L^{d}} \ {\phi}_{N,L}(A)\eqno(6.3)$$
and
$${\phi}(A[{\alpha}(t)B]C)={\rm lim}_{L\to\infty;N={\rho}L^{d}} \ 
{\phi}_{N,L}(A[{\alpha}_{N,L}(t)B]C)
\eqno(6.3)$$
for all local bounded observables $A,B,C$.
It is clear that the above formulae imply that ${\phi}$ does 
indeed inherit the ground state property of ${\phi}_{N,L}$, since the condition for this is 
that the support of the distribution valued Fourier transform of the function 
$t({\in}{\bf R}){\rightarrow}{\phi}(A{\alpha}_{t}B)$ lies in ${\bf R}_{+}$: as noted in 
Section 2, this characterisation of ground states is equivalent to the positivity condition 
on the Hamiltonian operator $H$ of Eq. (2.15).
\vskip 0.2cm
Turning now to space translations and Galilei boosts of ${\Sigma}_{N,L}$, we note that, 
in view 
of the periodicity of ${\Lambda}_{L}$, these are represented by the automorphisms 
${\sigma}_{N,L}(x)$ and ${\xi}_{N,L}(v_{L})$ of ${\cal A}_{L}$, respectively, given 
by the natural 
counterparts of Eqs. (2.6) and (2.7), though with the components of $v_{L}$ restricted to 
integral multiples of $2{\pi}/L$. Thus, the boosted ground state of ${\Sigma}_{N,L}$ is 
${\phi}_{N,L}{\circ}{\xi}_{N,L}(v_{L})$. To relate this to the boosted ground state 
${\phi}_{v}$ 
of ${\Sigma}$, we assume that
$${\rm lim}_{L\to\infty;N={\rho}L^{d}:v_{L}{\rightarrow}v} \ 
{\phi}_{N,L}({\xi}_{N,L}(v_{L})A)={\phi}_{v}(A)\eqno(6.5)$$
for all local observables $A$.
By Eqs. (6.5) and (2.7), the dynamical group associated to $\phi_{v}$ corresponds to a 
finite region Hamiltonian (2.11) given by
$$H_{\Lambda}= H^{v}_{\Lambda}= H_{N,L}+v.P_{N,L} \eqno(6.6)$$
We have subtracted from the Hamiltonian the constant term $N v^2/2$, which does not 
change the dynamical automorphism group. We remark that, in contrast to the situation 
expected, and sometimes proved, in relativistic quantum field theory [54], 
$H^{v}_{\Lambda}$ is not a positive operator.
Further, the space translational automorphisms ${\sigma}_{N,L}$ 
are implemented by a one parameter unitary group, whose generator is $i$ times the 
momentum 
operator $P_{N,L}$, i.e
$${\sigma}_{N,L}(x)A={\rm exp}(iP_{N,L}.x)A{\rm exp}(-iP_{N,L}.x).\eqno(6.7)$$
Assuming ${\phi}_{N,L}$ to be translationally invariant, it then follows that 
$P_{N,L}{\Phi}_{N,L}=0$.    
\vskip 0.2cm
We now introduce the elementary excitations of ${\Sigma}_{N,L}$ and then pass to 
those of ${\Sigma}$ in the following way. We assume that each of 
the elementary excitations of ${\Sigma}_{N,L}$  has a well-defined momentum $k$ and 
energy ${\varepsilon}(k)$ in the following sense. For each finite set $(k_{1},. \ k_{m})$ 
in ${\bf R}^{d}$, with $m=O(1)$ w.tr.t. $N$, there is a well-defined simultaneous 
eigenvector ${\Phi}_{N,L;k_{1},. \ .,k_{m}}$, of $H_{N,L}$ and $P_{N,L}$, such that 
$$P_{N,L}{\Phi}_{N,L}=(k_{1}+ \ +k_{m}){\Phi}_{N,L;k_{1},. \ .,k_{m}}
\eqno(6.8)$$
and
$$H_{N,L}{\Phi}_{N,L;k_{1},. \ .,k_{m}}=E_{N,L:k_{1},. \ ,k_{m}}
{\Phi}_{N,L;k_{1},. \ ,k_{m}}, \eqno(6.9a)$$
with
$$E_{N,L;k_{1},. \ ,k_{m}}=E_{N,L;G}+{\varepsilon}_{L}(k_{1})+ \ldots 
+{\varepsilon}_{L}(k_{m})+O(N^{-1}),\eqno(6.9b)$$
where $E_{N,L;G}$ is the ground state energy of ${\Sigma}_{N,L}$ .
This assumption is an abstraction of Lieb's results [49] on the Lieb-Liniger model [24]. 
There two branches of elementary excitations arise, and a choice must be made [49]. The 
term $O(N^{-1})$ is due to interactions between the elementary excitations: the latter 
occurs also in the case of the spin-waves in the Heisenberg model [55].
\vskip 0.2cm
As a first step to passing  from the excitations of ${\Sigma}_{N,L}$ to those of 
${\Sigma}$, we introduce the L. Schwartz space ${\cal D}(X^{m})$ and define 
$${\Phi}_{N,L;f }=(2{\pi}/L)^{md/2}{\sum}_{k_{1},. \ k_{m}}f(k_{1}, \ .,k_{m})
{\Phi}_{N,L;k_{1},. \ .,k_{m}} \ {\forall} \ f{\in}{\cal D}(X^{m}),\eqno(6.10)$$
where the $k$\rq s run over the integral multiples of $2{\pi}/L$. The factor 
$(2{\pi}/L)^{md/2}$ 
ensures that ${\Vert}{\Phi}_{N,L;f}{\Vert}$ reduces to a finite quantity, namely 
${\Vert}f{\Vert}_{L^{2}(X)}$, in the limit $N{\rightarrow}{\infty}$. We assume that 
the inner products of ${\Phi}_{N,L;f}$ with $[{\alpha}_{N,L}(t)A] {\Phi}_{N,L}, \ 
[{\sigma}_{N,L}(x)A]{\Phi}_{N,L}$ and $[{\xi}_{N,L}(v_{L})A]{\Phi}_{N,L}$ 
converge to 
canonical counterparts for the system ${\Sigma}$ as $N,L{\rightarrow}{\infty}, \ 
v_{L}{\rightarrow}v$ and $N/L^{d}={\rho}$. Thus, they serve to define a vector 
${\Phi}_{f}$ in 
the representation space ${\cal H}$ of ${\Sigma}$ with the properties that
$$({\Phi}_{f},[{\alpha}_{t}A]{\Phi})=
{\rm lim}_{L\to\infty;N={\rho}L^{d}} \ 
({\Phi}_{N,L;f},{\alpha}_{N,L}(t)A]{\Phi}),\eqno(6.11)$$
$$({\Phi}_{f},[{\sigma}(x)A]{\Phi})=
{\rm lim}_{L\to\infty;N={\rho}L^{d}} \ ({\Phi}_{N,L;f},{\sigma}_{N,L}(x)A]{\Phi}),
\eqno(6.12)$$
and
$$({\Phi}_{f},[{\xi}(v)A]{\Phi})=
{\rm lim}_{L\to\infty;N={\rho}L^{d};v_{L}{\rightarrow}v} \ 
({\Phi}_{N,L;f},{\xi}_{N,L}(v_{L})A]{\Phi}).
\eqno(6.13)$$
The vector ${\Phi}_{f}$ may now be unsmeared and expressed in terms of a vector 
valued 
distribution ${\Phi}_{k_{1},. \ .,k_{m}}$ according to the formula
$${\Phi}_{f}=\int_{X^{m}}dk_{1}. \ .dk_{m}f(k_{1},. \ .,k_{m}){\Phi}_{k_{1},. \ 
.,k_{m}}.\eqno(6.14)$$
It now follows from our specifications, especially the unitary implementations of the 
space and time translational automorphisms defined by Eqs. (2.14) and (2.16), that
$$U(t){\Phi}_{k_{1},. \ .,k_{m}}={\rm exp}\bigl(i({\varepsilon}(k_{1})+ \ 
+{\varepsilon}(k_{m}))t\bigr){\Phi}_{k_{1},. \ .,k_{m}}\eqno(6.15)$$
and 
$$S(x){\Phi}_{k_{1},. \ ,k_{m}}={\rm exp}\bigl(i(k_{1}+ \ +k_{m}).x\bigr)
{\Phi}_{k_{1}, \ ,k_{m}}.\eqno(6.16)$$
Hence, by Eqs. (2.15) and (2.17), ${\Phi}_{k_{1},. \ .,k_{m}}$ is a simultaneous 
eigenvector of the Hamiltonian, $H$, and momentum operator, $P$, such that
$$H{\Phi}_{k_{1},. \  ,k_{m}}=\bigl({\varepsilon}(k_{1})+ \ 
+{\varepsilon}(k_{m})\bigr)
{\Phi}_{k_{1}, \ .,k_{m}}\eqno(6.17)$$
and
$$P{\Phi}_{k_{1},. \ ,k_{m}}=(k_{1}+ \ +k_{m}){\Phi}_{k_{1},. \ .,k_{m}}.
\eqno(6.18)$$
These last two formulae represent the infinite volume limits of Eqs. (6.8) and 
(6.9).
\vskip 0.2cm
At this stage we are able to state what we mean by the generalised Landau state.
\vskip 0.3cm
{\bf Definition 6.1.} We say that ${\phi}_{v}$ is a {\it generalised Landau state} if the 
following condition - the {\it Landau superfluidity condition}- holds: there exists 
$v_{c}> 0$ such that
$${\varepsilon}(k)+v.k{\geq}0 \ {\forall} \ k{\in}{\bf R}^{d}
 \ {\rm if} \ {\vert}v{\vert}{\leq} v_{c}.\eqno(6.19)$$
Choosing $k$ in the direction of $-v$ in Eq. (6.19), we arrive at the well-known
explicit formula for $v_{c}$:
$$v_{c}= {\rm inf}_{k}({\varepsilon}(k)/{\vert}k{\vert})
\eqno(6.20)$$
By Eqs. (6.6), (6.8), (6.9), (6.17) and (6.18), Eq. (6.19) represents the stability of the state 
$\phi_{v}$ against generation of a finite number of elementary excitations. It should be 
emphasized that Landau formulated condition (6.19) at a not entirely quantum level, but 
rather in a semi-classical hydrodynamical framework. Sometimes the condition (6.19) is 
expressed, rather, in terms of quasi-particles. However, the meaning of this concept is 
somewhat vague, though Lieb [49] attempted to define it precisely as a pole of a Green's 
function. By contrast, the elementary excitations on which we base our considerations are 
exact eigenfunctions of the system satisfying certain well defined conditions [49].
\vskip 0.2cm
The sound velocity, $v_{s}$, is defined in terms of the elementary excitations by the 
formula
$$v_{s}={\vert}\bigl({\rm lim}_{k{\rightarrow}0}
{\partial}{\varepsilon}(k)/{\partial}k\bigr){\vert},\eqno(6.21)$$
On the other hand, by a macroscopic argument [2], one expects that $v_{s}$, is also 
given by
$$v_{s} = ({\rho}{\kappa}_{0})^{-1/2},\eqno(6.22)$$ 
where ${\kappa}_{0}$ is the compressibility of the system at zero temperature. This is 
defined in terms of the thermodynamic limit, $e({\rho})$, of the energy density and the 
pressure $P({\rho})$ by the formulae 
$$\kappa_{0}=\bigl[{\rho}{dP\over d{\rho}}\bigr]^{-1},\eqno(6.22)$$
where (cf. [19, P.58])
$$P={\rho}{de({\rho})\over d{\rho}}-e({\rho}).\eqno(6.23)$$
and
$$e({\rho})={\rm lim}_{L\to\infty;N={\rho}L^{d}} \ L^{-d}E_{N,L:G}.\eqno(6.24)$$
Under quite general conditions on the pair interactions, it may be proved that [32]
$${\kappa}_{0} \ {\geq} \ 0.\eqno(6.25)$$
Further it follows from Eqs. (6.19)-(6.21) that
$$v_{c}{\leq}v_{s}>0.\eqno(6.26)$$
We remark that, in the case of liquid Helium II,  heuristic, empirically supported 
arguments indicate that $v_{c}$ is less than $v_{s}$ and that its value is  given by the 
slope of the line through the origin in the ${\varepsilon}-{\vert}k{\vert}$ plane that is 
tangent to the graph of ${\varepsilon}$ in the neighbourhood of its local minimum (at 
$k{\neq}0$), which is governed by the roton spectrum [50]. 
\vskip 0.2cm
Clearly, a rigorous proof of the identity between the two
expressions (6.20) and (6.21) may be expected to be highly nontrivial. We shall come 
back to this point in section 6.3, where we prove it in a self-contained
manner for the Girardeau [25] model.
\vskip 0.3cm
{\bf 6.2. Stability condition for non- translationally covariant systems.}
The treatment of section 6.1 is only appropriate for {\it homogeneous} systems, in which 
momentum is conserved. The question may be posed, however, whether a natural 
replacement of the Landau superfluidity condition ( Definition 6.1) exists for 
inhomogeneous systems, such as the rotating system of Section 4 and the  LSY model of 
a dilute, trapped Bose gases [4, 18]. In this connection it is noteworthy that, according to  
the condition for translational superfluidity of the latter model, the free Bose gas is also a 
superfluid, while it is not so according to the Landau criterion (6.19). In this section we 
propose a replacement for the {\it necessity} part, represented by Eq. (6.26), of Landau's 
condition for inhomogeneous systems. Since  that inequality, together with condition 
(6.22),  signifies that the compressibility is finite, we proceed as follows.
\vskip 0.3cm\noindent
{\bf Definition 6.2.} We take the necessity part of Landau's
superfluidity condition, even for inhomogeneous systems, to be that the compressibility is 
finite, i.e. that
$${\kappa}_{0}<{\infty}.\eqno(6.27)$$
\vskip 0.3cm
For non-translationally invariant systems, the elementary excitations are eigenfunctions 
of  the Hamiltonian which do not carry a definite momentum, but show up in various 
quantities, such as the density of states and in the specific heat, displaying 
the characteristic phonon-behaviour $O(T^3)$ for low temperatures $T$. Since we are 
not able to formulate a stability condition in a mathematically precise sense as in Section 
6.1, we regard the word  ``superfluidity'' in Def. 6.2 in the phenomenological sense which 
is briefly discussed in Section 2, with $\rho_{n}=0$ in Eqs.(2.25)-(2.27) since we are 
referring to the ground state.
\vskip 0.2cm
We remark that,  in the case of the ideal Bose gas, the condition (6.27) is violated at 
$T=0$. It is not known rigorously as yet whether it holds there for dilute trapped bose 
gases in the Gross-Pitaevski limit, because the presently known rigorous bounds [4] do 
not allow  control over the compressibility.
\vskip 0.2cm
We devote the next section 6.3 to the Girardeau model [25], a special case of the Lieb-
Liniger model [24] ( the forthcoming Eq. (6.28)) which seems to be the only model for 
which the elementary excitations may be derived without further assumptions [49] and, 
moreover, allows an elementary derivation of the thermodynamic limit of the 
compressibility. The purpose of that section, is two-fold. On the one hand,the model 
allows an elementary proof of the identity between Eqs. (6.21) and (6.22), which is used 
to motivate definition (6.2); on the other, it shows very clearly how the repulsive 
interaction introduces a special structure of the ground state wave-function (quite 
different from a product of one-particle wave-functions in the same state, as in the free 
Bose gas), which complies with both definitions (6.1) and (6.2).
\vskip 0.3cm
{\bf 6.3. The Girardeau Model and the Landau superfluidity condition.} We start with the 
model of $N$ particles in one dimension with (repulsive) delta function interactions 
[24], whose formal Hamiltonian is given by
$$H_{N} = -{1\over 2}{\sum}_{i=1}^{N}{{\partial}^{2}\over {\partial}x^{2}}
+2c{\sum}_{i,j=1}^{{\prime}N}{\delta}(x_{i}-x_{j}), \ 0{\leq}x_{i}, \ 
x_{j}{\leq}L,\eqno(6.28)$$
the prime over the second ${\Sigma}$ indicating that summation is confined to nearest 
neighbours. For the (standard) rigorous definition corresponding
to (6.28) see [56].
\vskip 0.2cm
The limit as $c{\rightarrow}{\infty}$, of Eq. (6.28) yields Girardeau's model [25]. This 
seems to be the only model for which both the (double) spectrum of elementary 
excitations and the compressibility were obtained rigorously without (as yet) unproved 
assumptions.  The former were obtained by Lieb[49] and although the result for the 
compressibility follows from specializing the result of the appendix to that article (made 
rigorous by Dorlas [56]) to the present case,  the following derivation, which does not 
seem to be found in the literature, is elementary and provides a transparent physical 
reason for the validity of property (6.27). 
\vskip 0.2cm 
In the limit $c{\rightarrow}{\infty}$ of Eq. (6.28), the boundary condition on the wave-
functions reduces to 
$${\psi}(x_{1},. \ .,x_{N})=0 \ {\rm if} \ x_{j}=x_{l}, \ 1{\leq}j<l{\leq}N
\eqno(6.29)$$  
and the (Bose) eigenfunctions satisfying Eq. (6.29) simplify to
$${\psi}^{B}(x_{1},. \ .,x_{N})={\psi}^{F}(x_{1},. \ .,x_{N})
A(x_{1},. \ .,x_{N}),\eqno(6.30)$$
where ${\psi}^{F}$ is the Fermi wave function for the free system of
$N$ particles confined to the region $0{\leq}x_{i} < L, \ i=1,, \ .,N$,
with periodic boundary conditions, and
$$A(x_{1},. \ .,x_{N})={\prod}_{j<l}sgn(x_{j}-x_{l}).\eqno(6.31)$$
Note that ${\psi}^{F}$ automatically satisfies Eq. (6.29) by the exclusion principle. 
Indicating the Bose and Fermi ground states by the subscript $0$, it follows from 
Eq.(6.30) and the non-negativity of ${\psi}_{0}^{B}$ that
$${\psi}_{0}^{B} ={\vert}{\psi}_{0}^{F}{\vert}.\eqno(6.32)$$  
Since $A^{2}=1$, by Eq. (6.31), the correspondence between ${\psi}^{B}$ and 
${\psi}^{F}$ given by Eq. (6.30) preserves all scalar products, and therefore the energy 
spectrum of the Bose system is the same as of the free Fermi gas
Further, ${\psi}_{0}^{F}$ is a Slater determinant of plane-wave functions
labelled by wave-vectors $k_{i}, \ i=1,. \ .,N$, equally spaced over the range $[-
k_{F},k_{F}]$, where $k_{F}$, the Fermi momentum. Hence $k_{F}$ is equal to 
${\pi}(N-1)/L$, which in the thermodynamic limit reduces to 
$$k_{F}={\pi}{\rho}.\eqno(6.33)$$ 
The simplest excitation is obtained by moving a particle from $k_{F}$ to $q>k_{F}$ (or 
from $-k_{F}$ to $q< -k_{F}$), thereby leaving a hole at $k_{F}$ (or $-k_{F}$). This 
excitation
has momentum $p=(q-k_{F})$ (or $-(q-k_{F})$) and energy ${\varepsilon}(p)=(q^{2}-
k_{F}^{2})/2$, i.e.
$${\varepsilon}(p)=p^{2}/2+k_{F}{\vert}p{\vert}.\eqno(6.34)$$
It follows from this formula and Eqs. (6.21) and (6.34) that the sound velocity is
$$v_{s}= k_{F} ={\pi}{\rho}\eqno(6.35)$$
\vskip 0.2cm. 
On the other hand, the ground state energy density is given by the following standard 
formula for that of a one-dimensional ideal spinless Fermi gas in the thermodynamical 
limit (cf. [57, Sec. 56]).
$$e({\rho})=(2{\pi})^{-1}\int_{-k_{F}}^{k_{F}}dkk^{2}/2,$$
i.e. by Eq. (6.33),
$$e({\rho})={{\pi}^{2}\over 6}{\rho}^{3}\eqno(6.36)$$
and consequently, by Eqs. (6.23) and (6.24),
$$P={{\pi}^{2}\over 3}{\rho}^{3}\eqno(6.37)$$  
and
$${\kappa}_{0}^{-1}={\pi}^{2}{\rho}^{3}.\eqno(6.38)$$
According to this formula and Eq. (6.22), $v_{s}$ is given by Eq. (6.35), which confirms 
that, for Girardeau\rq s model, the phenomenological formula (6.22) yields the same 
result as the quantum mechanical one, (6.21). 
\vskip 0.2cm
We note that here the role of the repulsive interaction in leading to the Fermi distribution 
of momenta and thus to the finite compressibility. Equivalently, the structure, given by 
Eq. (6.32), of the ground state wave-function leads to a non-zero range of values of the 
particle momenta and hence to a non-zero velocity of sound. By contrast, for an ideal 
Bose gas, the particle momenta would all be concentrated at the value zero and, as a 
consequence, the speed of sound would be zero. 
\vskip 0.2cm 
In spite of its apparent simplicity, the above model is far from trivial. Some insight into 
its complexities is obtained when looking at correlation functions:
there exists a result due to Lenard [58] on the Fourier transform of the one-particle 
density matrix, implying the absence of ODLRO in the model, and just a few results on 
higher order correlations [59], restricted, however, to Dirichlet and Neumann boundary 
conditions.
\vskip 0.2cm
We conclude that the Landau condition depends very strongly on the specific structure of 
the ground state wave function, in contrast to London rigidity, which depends only on 
very general properties of the system, such as ODLRO. The latter is, however, not 
necessary for translational superfluidity in the Landau picture as shown, again, by the 
Girardeau model, bearing in mind Lenard\rq s above cited result.
\vskip 0.5cm
\centerline {\bf 7. Conclusion.}    
\vskip 0.3cm
In this paper we have analysed two aspects of the mathematical theory of superfluidity. 
These are centred on the ODLRO condition and on a precise form of Landau\rq s 
condition for the stability of uniform currents against elementary excitations. 
\vskip 0.2cm
In fact, the condition of ODLRO corresponds to that of gauge symmetry breakdown for 
the description of the bosonic model in terms of its field, rather than observable, algebra.
Thus, the equilibrium states with ODLRO enjoy the properties of both gauge and Galilei 
symmetry breakdown and the resultant emergence of Goldstone bosons and long range 
correlations are described in Section 3. Further, as shown in Section 4, the combination 
of ODLRO and axial symmetry leads to a  "rotational superfluidity" of the system, which 
is manifested when the system is placed in a rotation bucket. This kind of superfluidity 
stems a superselection rule that represents a macroscopic form of London rigidity. It is 
found to prevail in both the ideal Bose gas [51] and  in the LSY model of a dilute, trapped 
interacting Bose gas when the angular velocity of rotation ${\omega}$ does not exceed a 
critical value ${\omega}_{c}$. In the latter model, the Hamiltonian in the rotating frame 
becomes unbounded from below, due to an instability against the creation of an unlimited 
number of vortices, when  ${\omega} > {\omega}_{c}$ [43].  
\vskip 0.2cm
We have proved in Section 5 (Theorem 5.1) that, under very general conditions,  
translationally invariant current carrying states are not locally thermodynamically stable 
(LTS), in the sense specified in [38, 39]. This implies that a weaker kind of stability is 
needed to characterise the observed metastability of current carrying states. The picture 
of this metastability formulated in Section 6 is expressed in terms of the concept of  
generalised Landau states (Defs. 6.1 and 6.2), which are translationally invariant current 
carrying ones that are stable against finite numbers of {\it elementary excitations}, which 
in general are not the same as quasi-particles. This picture is realised by the Lieb-Liniger 
model [24] , under the assumption that the elementary excitations are given by the Bethe 
Ansatz. In the special case where this model reduces to that of Girardeau [25], no such 
supplementary assumption is needed, and our explict review of its properties 
demonstrates how this neo-Landau kind of superfluidity , by contrast with the rotational 
version,  depends very strongly on the special structural properties of the ground state.
\vskip 0.2cm
Finally, we should like to mention some open problems. The first is whether the necessity 
part of Landau\rq s condition (Def.  6.2) is realised by the model of dilute trapped gases 
[4]: the existing estimates do not, so far, allow control over the compressibility [4]. 
Another problem is whether Theorem 5.1 can be extended to a proof  that generalised 
Landau states do not satisfy the KMS or ground state conditions: what is lacking there is 
a proof that the LTS and the KMS (or ground state) conditions are equivalent for 
continuous systems, as they have been proved to be [38, 39] for lattice models. Evidently, 
that is a major problem, which goes beyond the context of the theory of superfluidity. So 
too is the problem of whether ground states on the algebra of observables extend to ones 
on the field algebra in the same way as KMS states have been proved [40] to do: a proof 
that they do so would establish a one-to-one correspondence between ODLRO and gauge 
symmetry breakdown and thereby complete the picture discussed in Section 3. 
\vskip 0.5cm\noindent
{\bf Acknowledgements.} This work was performed while we were
guests of the ESI, Vienna. We should like to thank Professor Jacob Yngvason for the 
invitation to ESI and to thanks the Staff of the Institute, as well as Professor Heide 
Narnhofer, for their kind hospitality.
\vskip 0.5cm
\centerline {\bf References}
\vskip 0.3cm\noindent
[1]  F. London: {\it Superfluids, Vol. 1}, Wiley, New York and Chapman and Hall, 
London, 1950
\vskip 0.2cm\noindent
[2] F. London: {\it Superfluids, Vol. 2},  Wiley, New York and Chapman and Hall, 
London, 1954
\vskip 0.2cm\noindent
[3] I. M. Khalatnikov: {\it Introduction to the Theory of Superfluidity}, Benjamin, 
New York, Amsterdam, 1965
\vskip 0.2cm\noindent
[4] E. H. Lieb, R. Seiringer, J. P.Solovej and J. Yangvason: {\it The Mathematics of 
the Bose gas and its Condensation}, Birkhaeuser, Basel, 2005
\vskip 0.2cm\noindent
[5] P. W. Higgs: Phys. Rev. {\bf 145}, 1156, 1966
\vskip 0.2cm\noindent
[6] R. Graham and H. Haken: Z. Phys. {\bf 237}, 31, 1970
\vskip 0.2cm\noindent
[7] H. Froehlich: Int. J. Quantum Chem. {\bf 2}, 641, 1968
\vskip 0.2cm\noindent
[8] L.Tisza: Phys. Rev. {\bf 72}, 838, 1947
\vskip 0.2cm\noindent
[9] O. Penrose and L. Onsager: {\it Bose-Einstein Condensation and Liquid Helium}, 
Phys. Rev. {\bf 104}, 576, 1956
\vskip 0.2cm\noindent
[10] C. N. Yang: Rev. Mod. Phys. {\bf 34}, 694, 1962
\vskip 0.2cm\noindent
[11] J. Bardeen, L. N. Cooper and J. R. Schrieffer: Phys. Rev. {\bf 108}, 1175, 1957
\vskip 0.2cm\noindent
[12] L.D. Landau: J. Phys. U.S.S.R. {\bf 5}, 71, 1941
\vskip 0.2cm\noindent
[13] N.N. Bogoliubov: J. Phys. (U.S.S.R.) {\bf 11}, 23, 1947.
\vskip 0.2cm\noindent
[14] V. Zagrebnov and J. B. Bru: Phys. Rep. {\bf 350}, 291, 2001
\vskip 0.2cm\noindent
[15]  R. P. Feynman:  in {\it Progress in Low Temperature Physics, Vol. 1}, Ed. C. J. 
Gorter, North Holland, Amsterdam, 1955
\vskip 0.2cm\noindent
[16]  G. L. Sewell: J. Math. Phys. {\bf 38}, 2053, 1997
\vskip 0.2cm\noindent
[17] E.H. Lieb and R. Seiringer: Phys. Rev. Lett. {\bf 88}, 170409-1-4, 2002
\vskip 0.2cm\noindent
[18] E. H. Lieb, R. Seiringer and J. Yngvason: Phys. Rev. B {\bf 66}, 134529, 2002
\vskip 0.2cm\noindent
[19] D. Ruelle: {\it Statisticsl Mechanics}, W. A. Benjamin, New York, 1969.
\vskip 0.2cm\noindent
[20] G. G. Emch: {\it Algebraic Methods in Statistical Mechanics and Quantum Field 
Theory}, Wiley, New York, 1972 
\vskip 0.2cm\noindent
[21] W. Thirring: {\it Quantum Mechanics of large Systems}, Springer, New York, 
Vienna, 1980 
\vskip 0.2cm\noindent
[22] G. L. Sewell: {\it Quantum Mechanics and its Emergent Macrophysics}, Princeton 
Univ. Press, Princeton, 2002.
\vskip 0.2cm\noindent
[23] L.Onsager: Nuov. Cim. {\bf 6}, 249, 1947
\vskip 0.2cm\noindent
[24] E. H. Lieb and W. Liniger: Phys. Rev. {\bf 130}, 1605, 1963
\vskip 0.2cm\noindent
[25] M. D. Girardeau: J. Math. Phys. {\bf 1}, 516, 1960
\vskip 0.2cm\noindent
[26] G. F. Dell'Antonio, S. Doplicher and D. Ruelle: Commun. Math. Phys. {\bf 2}, 
223, 1966
\vskip 0.2cm\noindent
[27] G. L. Sewell: J. Math. Phys. {\bf 11}, 1868, 1970
\vskip 0.2cm\noindent
[28] S. Miracle-Sole and D. W. Robinson: Commun. Math. Phys. {\bf 14}, 235, 1969
\vskip 0.2cm\noindent
[29] D. A. Dubin  and G. L. Sewell: J. Math. Phys. {\bf 11}, 2990, 1970
\vskip 0.2cm\noindent
[30] C. Radin: Commun. Math. Phys. {\bf 54}, 69, 1977
\vskip 0.2cm\noindent
[31] G. L. Sewell: Lett. Math. Phys. {\bf  6}, 209, 1982
\vskip 0.2cm\noindent
[32] M.E.Fisher: Arch. Rat. Mech. Anal. {\bf 17}, 377, 1964
\vskip 0.2cm\noindent
[33]  I. E. Segal: Ann. Math. {\bf 48}, 930, 1947
\vskip 0.2cm\noindent
[34] R. Haag, N. M. Hugenholtz and M. Winnink: Commun. Math. Phys. {\bf 5}, 
215, 1967
\vskip 0.2cm\noindent
[35] G. G. Emch, H. J. F. Knops and E. Verboven: J. Math. Phys. {\bf 11}, 1655, 1970 
\vskip 0.2cm\noindent
[36] R. Haag, D. Kastler and E. B. Trych-Pohlmeyer: Commun. Math. Phys. 
{\bf  38}, 173, 1974
\vskip 0.2cm\noindent
[37]  W. Pusz and S. Woronowicz, Commun. Math. Phys. {\bf 58}, 273, 1978
\vskip 0.2cm\noindent
[38] H. Araki and G. L. Sewell: Commun. Math. Phys. {\bf 52}, 103, 1977
\vskip 0.2cm\noindent
[39] G. L. Sewell: Phys. Rep. {\bf 57}, 307, 1980
\vskip 0.2cm\noindent
[40] H. Araki, R. Haag, D. Kastler and M. Takesaki: Commun. Math. Phys. 
{\bf  53,}, 97, 1977
\vskip 0.2cm\noindent
[41] E. P. Gross: J. Math. Phys. {\bf 4}, 195, 1963
\vskip 0.2cm\noindent
[42] L. P. Pitaevski: Sov. Phys. JETP {\bf 13}, 451, 1961
\vskip 0.2cm\noindent
[43] R. Seiringer: J. Phys. A {\bf 36}, 9755, 2003
\vskip 0.2cm\noindent
[44] J. A. Swieca: Commun. Math. Phys. {\bf 4}, 1, 1967
\vskip 0.2cm\noindent
[45] L. J. Landau,  J. F. Perez and W. F. Wreszinski: J. Stat. Phys. {\bf 26}, 755, 1981
\vskip 0.2cm\noindent
[46]  D. Ruelle: {\it Symmetry Breakdown in Statistical Mechanics}, Pp. 169-194 of {\it  
Cargese Lectures, Vol. 4}, Ed. D. Kastler, Gordon and Breach, New York, 1969   
\vskip 0.2cm\noindent
[47] M. Fannes, J.Pule and A. Verbeure: Lett. Math. Phys. {\bf 6}, 385, 1982 
\vskip 0.2cm\noindent
[48] W. F. Wreszinski: Forts. Phys. {\bf 35}, 379, 1987
\vskip 0.2cm\noindent
[49] E. H. Lieb: Phys. Rev. {\bf 130}, 1616, 1963
\vskip 0.2cm\noindent
[50] R.P. Feynman: Statistical Mechanics- A Set of Lectures- Benjamin, 
N.Y.1972
\vskip 0.2cm\noindent
[51] J. T. Lewis and J. Pule: Commu. Math. Phys. {\bf 45}, 115, 1975.
\vskip 0.2cm\noindent
[52]  E. H. Lieb and R. Seiringer: Commun. Math. Phys. {\bf 264}, 505, 2006
\vskip 0.2cm\noindent
[53] W. F. Stinespring: Proc. Amer. Math. Soc. {\bf 6}, 211, 1955.
\vskip 0.2cm\noindent
[54] W. F. Wreszinski and C. Jaekel: Ann.Phys.(N.Y.) {\bf 323}, 251, 2008
\vskip 0.2cm\noindent
[55] K.Hepp: Phys. Rev. {\bf B5}, 95, 1972
\vskip 0.2cm\noindent
[56] T. C. Dorlas: Comm. Math. Phys. {\bf 154}, 347, 1993
\vskip 0.2cm\noindent
[57] L. D. Landau and E. M. Lifschitz: {\it Statistical Physics}, Pergamon, London, 
Paris, 1959  
[58] A. Lenard: J. Math.Phys. {\bf 5}, 930, 1964
\vskip 0.2cm\noindent
[59] P.J. Forrester, N. Frankel and Garoni, T M: J. Math. Phys. {\bf 44}, 4157,
2003
\vskip 0.2cm\noindent
\end